%% file: ms.tex
\documentstyle[12pt,aasms4]{article}
\input ./abbrevs
\begin{document}

\title{The Role of a Massive Central Singularity in Galactic 
Mergers on the Survival of the Core Fundamental Plane.} 
\author{Kelly Holley-Bockelmann and Douglas O. Richstone}
\affil{Astronomy Department, University of Michigan}

\lefthead{Holley-Bockelmann \& Richstone}
\righthead{Survival of the Core Fundamental Plane}

\authoraddr{Ann Arbor, MI 48109-1090}

\begin{abstract}

In order for the core Fundamental Plane (cFP) to endure, small
ellipticals must not survive mergers with giant ellipticals, despite
the fact the small secondary galaxy can be as much as a million times
more dense than the primary. However, our previous set of experiments has
shown that, for purely stellar galaxies, the secondary does in fact
survive mergers with primaries up to 100 times more massive. In this
paper, we demonstrate the effect that a massive central black hole has
on mergers of cFP galaxies.  Our results indicate that the addition
of a massive central singularity inside the primary galaxy provides
strong enough tidal forces to destroy dense cFP companions when the
secondary's orbital decay is sufficiently elongated. The 
destruction of the secondary acts to preserve the original low central 
density profile of the primary in the giant merger remnant, which allows 
the remnant to remain on the cFP. On more circular orbits, though, the 
secondary is only disrupted near the end of the merger, and the degree to 
which the secondary particles disperse depends on the amount of 
orbital energy left in the merger. Hence, there are some mergers for which
the cFP is not preserved in our experiments. We find that if the secondary 
is not dispersed, it forms a spinning stellar disk with a central density 
that forces the merger remnant off the cFP. 
 
\end{abstract}

\section{Introduction}

Although elliptical galaxies may evolve passively, there is considerable
evidence that at least some ellipticals evolve by merging. For example,
counterrotating cores can be explained by the recent accretion of a 
spinning galaxy, especially when accompanied by a secondary 
starburst (Kormendy, 1984; Franx \& Illingworth, 1988; Carollo 
{\it et al}, 1997). Mergers are also thought to be responsible 
for the apparent bimodality of globular cluster populations 
(Kissler-Patig {\it et al}, 1998; Whitmore, 
1997; Ashman \& Zepf, 1992), and for structural changes like multiple 
nucleii, dust lanes, circumnuclear shells, and boxy isophotes 
(Malin \& Carter, 1983; Schweizer, 1982; Seitzer \& Schweizer, 1990; 
Forbes \& Thomson, 1992).

To the extent that elliptical galaxies evolve by satellite accretion, 
it is difficult 
for a large, gas-poor elliptical galaxy to preserve a low density core, 
because the accretion of a small, high density 
satellite will steepen the inner density profile of the merger remnant 
(Faber {\it et al}, 1997), provided the secondary survives. This is the 
paradox of the core Fundamental 
Plane (cFP).  The cFP demonstrates that elliptical 
galaxy centers maintain a tight relationship between projected 
central density and luminosity, a relationship foreshadowed independently by 
Lauer and Kormendy 15 years ago (Kormendy, 1985; Lauer, 1985).
If, however, a secondary survives in a high density 
ratio merger (that is: if the central phase space distribution of the 
secondary and its debris has not dispersed compared to its original 
state), then the merger remnant will not lie on this plane. So, if even a 
fraction of large ellipticals accrete dense cFP secondaries, we would
expect considerable scatter at the bright end of the cFP.

A massive black hole at the center of large galaxies may 
preserve the cFP by tidally disrupting dense secondaries in 
the accretion process. Observations are beginning to
show that massive central black holes are a natural part of 
galaxy centers (Richstone {\it et al}, 1998), and as a consequence, their effects ought to be included
in calculations of galaxy mergers. Current dynamical estimates of the 
best galactic black hole candidates have yielded masses on the order of 
$0.005 M_{\rm bulge}$ (Magorrian {\it et al}, 1998; Kormendy \& Richstone, 1995). Such massive black holes dominate the galactic potential inside 
the cusp radius, $r_{\rm cusp} \approx G M_{\bullet}/ \sigma_{\rm bulge} ^2$,
where $M_{\bullet}$ is the black hole mass, and $\sigma_{\rm bulge}$ is the
velocity dispersion of the bulge.
This cusp radius can be on the order of a kiloparsec for the largest
ellipticals, which is not a small fraction of the core radius and 
can, in some cases, be resolved.  

Despite the importance of black holes to elliptical galaxy centers, 
simulating the effects of a massive black hole 
on the stellar distribution unfortunately presents a numerical challenge.
Stellar velocities increase as $r^{-1/2}$ near the black hole, and the
tiny timestep required to accurately track these stars is prohibitively
expensive. In addition, the steep gradient in the potential near a
black hole must be well resolved, and this typically requires 
a very large particle number (see however Sigurdsson {\it et al}, 1995).
Nonetheless, there has been significant work done both on developing 
realistic galaxy models with central singularities 
(Merritt \& Quinlan, 1998; Sigurdsson {\it et al}, 1995), and on simulating 
the effect that black hole binaries have on the host galaxy 
(Makino \& Ebisuzaki, 1996; Governato {\it et al}, 1994). In addition,
the analysis of an ensemble of individual stellar orbits within a
black hole embedded galaxy may indicate the destabilizing influence 
of a black hole on a galaxy's orbital structure (Merritt \& Valluri, 1998).

In a previous paper, we showed that for purely stellar cFP galaxies,
a secondary survives any merger in which it is more dense than
the primary (Holley-Bockelmann \& Richstone, 1999, hereafter paper 1). 
In this paper, we isolate the effect of a single massive black hole in these 
encounters by adding a black hole to the primary galaxy. As in paper 1, 
the primary is rigid, so the addition of a black hole 
can be modeled as a external force on the secondary. While the 
secondary is separated from the black hole (or if the encounter is impulsive), there is no need to invoke 
a tiny timestep or an increased particle number in the secondary. This 
approach is an efficient way to determine whether massive central black holes 
can destroy a dense satellite during a merger. In this paper, we apply the 
method developed to problem of secondary destruction. 
We begin with a review of our approximation method in $\S$ 2. For 
details of the technique, in particular for tests of both the rigid 
primary approximation and our particle-field code, please refer to paper 1. 
The tests of the black hole embedded method and results of our 
simulations can be found in $\S$ 3. Section 4 discusses the 
implications of the results on the persistence of the
cFP and previews future work.

\section{Methods}
\subsection{The Galaxy Models}

We used the same technique as in paper 1 for choosing 
initial conditions, and for defining, modeling and populating galaxies on 
the cFP. See Table 1 for galaxy parameters and black hole masses.
Our galaxies were initially composed of 5000 particles distributed 
over both the core and the envelope of our galaxies. For each mass ratio, 
a test merger was run, and the particle loss of the envelope was analyzed.  
To achieve better central resolution, we conducted the merger again, this 
time assigning 5000 particles to only the central
regions of the secondary under the assumption that the galaxy envelope
behaved in the same manner as in our test run. Table 2 presents the
spatial resolution for our double and single component galaxy models. 
We followed only the particles that were bound to the secondary, 
but we preserved the phase space, energy, and angular momentum information 
of the unbound particles at the time they were stripped from the secondary. 
We will concentrate on the better resolved results for the 5000 particle 
single component galaxies, also referred to as inner $\eta$ models.

\subsection{The Force on the Secondary}

Since the addition of a central black hole introduces a force that
does not tend smoothly to zero at the center of the primary,
we chose not to use a tidal approximation of the external force,
as we did in paper 1. Instead, we advanced the particles in the 
inertial frame, where the force on a secondary particle is:

\begin{equation} \vec F_{\rm tot}(\vec R) = \vec F_{\rm 2}(\vec r) + 
\vec F_{\rm {\rm 1}}(\vec R)^ {\star} +\vec F_{\rm fric}(\vec R) +
\vec F_{\rm {\bullet}}(\vec R), \end{equation}

\noindent
where $\vec F_{\rm 2}$ is the self gravity of the secondary,
$\vec F_{\rm {1}}^ {\star}$ is the force on a secondary particle due to the 
stars in the primary, $ \vec F_{\rm {fric}}$ is the force due to dynamical 
friction, $\vec r$
is the vector which points from the secondary center to a secondary particle, 
$\vec R$ is the vector which points from the primary center to the secondary
particle, and $\vec F_{\bullet}$ is the force due to the black hole, 
expressed as:

\begin{equation} \vec F_{\bullet}(\vec R) = - {{G M_{\bullet} m_{\rm p}} \over {(\vec R + \epsilon)^3}} \vec R, \end{equation}

where $\epsilon$ is a softening parameter which we chose to be close to 
the core radius of the
secondary, and $M_{\bullet}$ was chosen to be consistent with the Kormendy \&
Richstone relation (1995). Our softening parameter is much larger
than the spatial resolution in our inner $\eta$ models, but we chose
this larger $\epsilon$ because it was consistent with the spatial resolution
in the double $\eta$ model galaxies here and in paper 1, and we wanted to 
be sure that we were isolating the effect of the black hole when we compared
our results to our previous experiments. With this degree of softening,
the experiments are designed to represent a lower limit to the damage done
to dense secondaries. 

\subsection{The Dynamical Friction on a Secondary Particle}

We apportioned the total dynamical friction force, $F_{\rm {fric}}$,
equally to each secondary particle (see appendix A).
The frictional acceleration applied to each particle in the secondary galaxy, 
$ d \vec v_{f} / dt$, is derived from the Chandrasekhar 
formula and is a function of a secondary's position in the primary galaxy:

\begin{equation} {d \vec v_{f}(\vec R) \over dt} = - {f_{\rm drag} \> {{{4 \pi \ {\rm ln} \Lambda \ G^2 \rho_{1} \ {m_2(t)}}} \;\Bigl[ {\rm {erf}}(X) - {{{2X}\over{\sqrt{\pi}}}e^{-{X}^2}}\Bigr] \; {{\vec v_{2}}  \over { \ \vec {v_{2}}^3}}}}, \end{equation}

\noindent
where $\vec v_{2}$ is the velocity of the secondary, 
$X \equiv v_{2} / ( \sqrt{2} \sigma)$, $\sigma = \sqrt {0.4 G M_{1}/r_{1, {\rm eff}}}$, $\Lambda$ is the Coulomb logarithm which was set to $M_{1}/M_{2}$, and
$f_{\rm drag}$ is a drag coefficient, as explained in paper 1. We allowed
the total mass of the secondary galaxy, $m_{2}(t)$ to vary as mass is
lost in the merger.  

Mass lost by the secondary will decrease the magnitude of the dynamical
friction force, and will change the orbital decay
trajectory such that the secondary experiences more pericenter passes.
Analysis of the purely stellar simulations from paper 1 indicate that 
the secondary was stripped at each pericenter pass down to its tidal radius, 
$ {r_{\rm tide}}^ 3 \approx  {{M_{1}} / { M_{2}}} \ {D}^3 $. 
We incorporated this knowledge 
into our set of black hole simulations. We initially set the mass of the 
secondary in $a_{\rm {fric}}$ equal to the total secondary mass, and when 
the secondary encountered a pericenter pass that was within the core 
radius of the primary, we reset the total mass to the mass enclosed by
the tidal radius. \footnote{We did not include secondary mass loss in 
the orbital decay calculations in paper 1. Therefore, we simulated the 
purely stellar 10:1 merger from paper 1 again with an orbital trajectory that 
included mass loss in the dynamical friction term as described above.
Since the secondary remained intact, we can be certain that  
it was not the change in the orbital decay that destroys the 
secondary in our black hole experiments (figure 1). }

This two part mass loss scenario 
was selected for reasons of computational speed, since a continuous mass 
loss term in the orbital decay would result in a larger number of
orbits that are far from the damaging black hole potential. These large 
apocenter orbits take a long time to integrate, and most of the integration
is spent following a secondary that is too far from the center
of the primary to feel a significant external force. While this scheme 
underestimates the change in the dynamical friction force, 
resulting in fewer pericenter passes than in reality, it useful as a 
limiting experiment: if the secondary breaks up after a few pericenter 
passes, it would certainly break up after more.

\section{Results}

As in paper 1, we explore a 2-dimensional grid of parameter space.
In the first dimension, we vary the mass ratio, and in the second
dimension, we vary the initial orbital angular momentum of the secondary.
In this section, the results are organized in groups of differing 
angular momentum. Since our goal is to investigate the effect of a 
single massive black hole has on the breakup or survival of dense 
secondaries in mergers with primaries, first sought to duplicate the
plunging encounters we explored in paper 1. In this way, we can isolate
the damage generated by the black hole. For a summary of
the different experiments conducted, see table 3.  

\subsection{Plunging Orbits}

Anticipating that the most damaging effect would be due to tidal
forces experienced as the secondary passes through the center of the
primary, we launched a series of nearly parabolic, plunging orbits.
In this basic set of experiments, we investigated two mass ratios:
100:1, 10:1. These mass ratios correspond to density
ratios of approximately 1:830 and 1:105 at a radius of 0.1 pc, 
and the density ratio rapidly increases at smaller radii.
We focus on the 
10:1 simulation. Figure 2 shows the time evolution of the secondary
in the force field of the primary.  Here, we learn that most of the
mass is stripped at pericenter, and comes off impulsively in a cloud
which continues to expand as the secondary crosses the primary center
again. However, it is apparent that the impulsive injection of
energy into the secondary by the black hole is not enough to 
disrupt the secondary after the first pericenter pass. 
From the impulse approximation, the first order change in energy of 
the secondary due to the black hole,

\begin{equation} {\Delta E / E} \approx ({{M_{\bullet}}/{M_{2}}) \ ({{ \sigma_{2}}/{v_{\rm orb}}}})\ ({{R_{\rm core}}/{P}}), \end{equation}

\noindent is of order $10 ^{-2}$ on the first pass. 
Instead, the secondary remained 
intact inside the tidal radius, $ {r_{\rm tide}}^ 3 \approx  
{{M_{1}} / { M_{2}}} \ {S}^3 $, for approximately 5 more pericenter passes,
until enough energy was pumped into the secondary to unbind it. In figure 3, 
we illustrate the change in the secondary energy
from beginning to end for this merger. The stripped mass is still bound 
to the primary and is spread out over a volume of space with a radius of approximately 250 parsecs from the black hole, or roughly the second apocenter distance. With the secondary disrupted over so large a volume,
the merger remnant will remain on the cFP.  
Figure 4 illustrates the change in the secondary density profile for 
the 10:1 mass ratio. For the 100:1 mass ratio, the secondary density 
profile and snapshots of the secondary during the merger are presented in 
figures 5 and 6, respectively. To ensure that 
the secondary's disruption was not an error caused
by pericenter passes directly through the black hole, we also conducted a
$\kappa=0.05$, 100:1 experiment (figure 7), in which the first pericenter pass
was approximately 400 parsecs from the black hole, on the order of the core 
radius of the primary. In all three experiments,
the secondary was destroyed. This is a markedly different result
from the purely stellar case where the secondary remained intact,
and indicates the importance of massive black holes as a
source of impulsive energy during mergers.

\subsection{Small Black Hole Mass}

To confirm that the black hole is the direct cause of the secondary's
destruction, we launched a zero angular momentum 10:1 mass ratio 
encounter in which the primary was host to a black hole with
about $0.005 \%$ the mass of the secondary, or nearly $9 x 10^{6} M_{\odot}$.
In this case, the center region of the secondary was preserved (figure 8).
We interpret this result as evidence that our method can detect 
a secondary's disruption yet does not force disruption erroneously through
the simple existence of a singularity. It is conceivable that the cFP 
can put a constraint on the mass of central black holes, independent 
of AGN light predictions; clearly, when the black hole mass is down
by a factor of over 300 from the ridgeline, as it is in this experiment, the 
mass is not sufficient to destroy a secondary.  Faber {\it et al}, 1997
presents a similar argument that the central black hole mass is constrained
by the mass of the stellar core profile that forms around a binary 
black hole pair.

\subsection{Non-radial Secondary Orbits}

\subsubsection{Secondary Destruction}

To explore the effect that different
orbits have on the preservation of the cFP, we launched
secondaries in the 10:1 and 100:1 mass ratios on orbits with significant 
angular momentum. Our orbits are parameterized by ${\kappa} \equiv {{L} / 
{L_{\rm circ}}}$. In the 10:1 mass ratio, we selected $\kappa = 0.2, 0.8$,
and for the 100:1 case, we chose $\kappa= 0.5$. Figure 9 displays an
xy plane projection of the first few orbits in each of these trajectories.

In each of these experiments, the secondary was stripped to
the tidal radius at the last pericenter of the the orbit. However, since 
the secondary
is far from the black hole on all but the last few pericenter passes,
this tidal radius is quite large for most of the decay trajectory.
The secondary's core is therefore intact through all but the last 
passes, since the density at the secondary core is clearly greater
than the stellar component of the primary everywhere. In the final passes,
the black hole can exert significant tidal forces on the
secondary, and the secondary is actually tidally compressed in two 
dimensions, which increases
its central density profile briefly in projection. 

On any given orbit, the secondary will eventually reach a place in the 
merger where the force exerted by the black hole is enough to overpower the
secondary self gravity and it is destroyed. Roughly speaking, this occurs the
first time the galaxy passes through a region where:

\begin{equation} {{M_{\bullet}} \over {{P}^3}} > {{M_2} \over {{r_2}^3}}, \end{equation}

\noindent where P is the distance of a pericenter pass, $M_{\bullet}$ is 
the black hole mass, and $r_2$ is the
core radius of the secondary. Strictly
speaking, all of our non-radial mergers resulted in the destruction of 
the secondary. However, it is insufficient to equate secondary damage with success in protecting the
cFP. While it is true that the core Fundamental Plane 
requires the disruption of dense secondaries in mergers, it is $not$ true 
that the mere disruption of secondary necessarily results in a remnant 
that lies on the cFP. At disruption, the particles inherit 
the orbital energy of the secondary, and for the particles to be dispersed,
the disruption must occur while there is still enough orbital energy to 
carry the debris out to a large apocenter so that the density of the debris 
is reduced to the density of the remnant. Otherwise, the remnant has
too steep a central density profile to remain on the cFP.
This can condition be stated as a function of the first post disruption 
apocenter:

\begin{equation} {{M_1} \over {{r_1}^3}} \approx {{M_2} \over {{A}^3}}, \end{equation}

\noindent where $r_1$ is the core radius of the primary and A is the first
post disruption apocenter. For the purposes of this paper, we 
define destruction to occur only when the debris is spread over
a large enough apocenter to result in a significantly lower density 
profile. Likewise, we define the disruptions that occur only when the merger 
is effectively complete to be survivals. 

Under this definition of destruction, the secondaries in the 10:1 $\kappa=0.2$ 
and $\kappa=0.8$ survived. See figures 10 and 11 for the change in the density
profile of the secondary for the 10:1, $\kappa = 0.2$ and 10:1, $\kappa=0.8$ 
experiments, respectively. The secondary in the 100:1, $\kappa=0.5$ merger, 
however, was destroyed. Figure 12 shows the 
density profile for the 100:1, $\kappa=0.5$ encounter.
In fact, from our derived destruction criterion, we predict that
for these galaxy parameters and our choice of dynamical friction, $all$
orbits in the 100:1 mass ratio will result in destruction.

We caution that the distance of the first post disruption apocenter
depends critically on the magnitude of the dynamical friction force at 
pericenter, which itself depends on the shape of the primary density 
profile and the mass of the secondary. For larger mass ratios (ie smaller
secondary mass), since the dynamical friction is weaker, the orbit retains
considerable orbital energy, and therefore experiences larger apocenters. 
Hence it is easier for the 100:1 case to 
disrupt than the 10:1 case. Similarly, a flat central primary density 
profile produces less dynamical friction at pericenter than a  
a primary with a steep central density. Therefore, an $\eta = 3.0 $ primary
are more likely to disrupt a secondary than an $\eta=1.0$ primary for
a given mass ratio.
With these
conclusions, we numerically integrated several orbits with various
mass ratios and primary density profiles and applied the 
disruption criteria in equation 5 (figure 13). 
From these results, we predict that if we were to
run our 10:1, $\kappa=0.5$ experiment with an $\eta=3.0$ primary, 	
the secondary would be $destroyed$, due mostly to the flatness
of the central primary density. 

Although our results are not general, in that they appear to 
depend on the choice of the primary, we note the following important result:
these experiments have produced a set of mergers that
were not destroyed by the addition of a massive central black hole.
Black holes, then, do not universally protect the cFP. Additionally,
we discovered that the preservation of the cFP depends critically
on the character of the orbit at the end of the merger.

\subsubsection{Disk Formation}

For non-radial mergers, the final part of the orbit produces interesting 
qualitative changes in the secondary as well. We define $P_2$ to be a 
pericenter pass that occurs before the secondary is disrupted. 
If $P_2$ is on the order of the
core radius of the secondary, as was the case
for the 10:1, $\kappa=0.2$ and $\kappa=0.8$ experiments, then
the secondary is tidally torqued into a spinning disk with the
radius of $P_2$ (orbital
angular momentum is transferred into spinning up the secondary).
A sense of the spin and a suggestion that the secondary is tidally torqued
in the $\kappa=0.8$ experiment
may be seen in figure 14.  
Figure 15 shows the flattening induced, in part, by the spin in 
the $\kappa=0.2$ encounter, and figure 16 illustrates the increase 
in the secondary spin for this experiment. Since the stellar disk begins 
to spin as it is still in the final orbits, it can be displaced from the 
primary center for a short time. For our $\kappa=0.2$ merger, the spinning 
stellar disk was detectably off center for approximately $7$ x $10^7$ years.
This is reminiscent of the off-center dust disk observed in NGC 4261 
(Ferrarese {\it et. al.}, 1996). At the end of the simulation, we
have a non self-gravitating central spinning stellar disk, which
has some resemblance to the stellar disk in NGC 3115 (Kormendy {\it et. al.}, 1996). However, with an aspect ratio of 4:1, the disk we have formed is much 
thicker than NGC 3115, which has an aspect ratio of 100:1. It is possible that
with the addition of gas dynamics to our simulations, energy loss through
gas dissipation could form a disk as thin as NGC 3115.

In the 10:1, $\kappa = 0.2$ and $\kappa = 0.8$ encounters, the high 
angular momentum particles were stripped preferentially from the
forming disk, and by the time disk reaches the primary center, 
mostly plunging orbits were left. These plunging orbits become unbound to 
the secondary when they pass close to the black hole, so the secondary 
is dissolved within a few crossing times of reaching the center (although
again, this denotes survival in our definition, because the debris
is as tightly bound to the remnant as it was to the secondary.). 
Figure 17 shows the energy/angular momentum distribution for the $\kappa=0.2 $
experiment.

\section{Discussion}
\vskip0.5cm

We have investigated the effect a massive black 
hole at the center of an otherwise purely stellar primary 
has on mergers of high density ratio galaxies on the cFP.
 
We have concluded that the amount of damage that the 
black hole can inflict on the secondary during a merger
is highly dependent on the orbital decay trajectory of the secondary. 
If the secondary's orbital decay is deeply plunging, the secondary
encounters the black hole potential impulsively, and
through repeated impulsive encounters, the black hole pumps
enough energy into the secondary to unbind it. 

If the secondary's encounter is non-radial, the secondary 
is far outside the radius of influence of the black hole for most 
of the merger, and it is stripped merely to the Roche radius. 
In our simulations, the damage done to the secondary center during this early 
stage is quite minimal. Only on
the final few orbits does the galaxy sink close enough to the 
black hole to experience significant tidal stripping, which
eventually unbinds the dense secondary center. However, unless the disruption
occurs while the merger has sufficient orbital energy, the debris 
orbits tightly
around the center of the primary, and the remnant density is increased.

An important feature of the merger trajectory is the dependence on the
density profile of the primary. In primaries with shallow density 
profiles and embedded black holes, the dynamical friction acting at 
the disruption is smaller than in primaries with steep density profiles, 
so the debris is more easily dispersed. For
these galaxies, a much wider range of secondary masses will destroyed
(that is: disrupted such that $\rho_1 \approx \rho_2$). 

It is tempting to identify this feature with the dichotomy between
the central light profiles of bright and faint galaxies. Faber \et 1997
note that galaxies brighter than about $ M_v = -21 $ have shallow, 
low density cores (their projected surface brightnesses 
$ -d log I / d log r < 0.5 $), while fainter ones are steeper. 
In our experiments, galaxies brighter than $M_v = -21.7$ destroy high mass ratio secondaries 
much more efficiently than fainter ones. Further investigation of
this result seems worthwhile.

A proper understanding of the preservation of the cFP requires knowledge
of the distribution of mass ratios and impact parameters of present day bulges.
This can perhaps be computed reliably in virialized clusters (Tormen, 1997),
where the galaxies encounter unbound targets. In this case, low-mass secondaries tend to merge on more circular orbits. The situation is likely to
be quite different in a cold Hubble flow, where progenitors encounter each
other only if they are on bound orbits, and where are the orbits are
likely to have little angular momentum (Aarseth \& Fall, 1980). It is hard to 
believe, however, that progenitors of bulges encounters each other with
{\it no} angular momentum, so it is not completely clear whether this
work indicates the true resolution to the paradox of the cFP, or whether
additional physics is needed in our experiments
to explain the persistence of the cFP. 

If $both$ the galaxies in a cFP merger have central black holes, then a black hole binary may form in the 
center of the merger remnant, and 3 body scattering may heat the 
center and lower the remnant's central density, allowing it to lie 
on the cFP. Apparently, there is some debate as to whether
black hole binaries form from high mass ratio mergers 
(Governato, \et, 1994). However, for equal mass mergers, Makino \& Ebisuzaki (1996), and Quinlan \& Hernquist (1997) 
found considerable black hole binary heating. 
As a consequence, the high central density from the
more circular encounters in this paper may be disrupted upon 
the introduction of a black hole in the secondary, as long as 
the black holes form a binary pair. If so, this may be a powerful case
that a black hole resides in the center of every galaxy. We will
present the results of the effects of multiple black holes in cFP mergers
in a future paper.

A second interesting feature of these results is the formation
of rapidly spinning disks. When the secondary is not destroyed on 
a non-radial encounter, it can begin to spin during the 
final orbits as it is torqued
by the black hole. Spinning stellar
disks have been observed in many galaxies (such as NGC 3115, Capaccioli \et, 1987), and have been invoked to explain apparent multiple nucleii in others 
(M31, Tremaine, 1995; NGC 4486b, Lauer \et, 1996).
Our purely stellar simulations form rather thick disks, with aspect ratios
of approximately 4:1. However, the formation of fat stellar disks seems inevitable in these mergers. To form a disk
as razor thin as NGC 3115 would most likely require a dissipative component. 
Nonetheless, non-radial galaxy mergers appear to be an efficient way to make
a spinning stellar disk, as long as one of the galaxies is embedded with
a massive central black hole.

Some support for this work was provided by 
the Space Telescope Science Institute, through general observer grant 
GO-06099.05-94A, and by NASA through a theory grant G-NAG5-2758.
We thank the members of the NUKER collaboration for helpful conversations. 
DR thanks the John Simon Guggenheim Foundation for a Fellowship.

\appendix
\section {Appendix A}

Under the impulse approximation, any single particle with mass m 
experiences a deflection, $\Theta$, when traveling with a velocity $\vec v$ past a single 
massive particle of mass M as follows:

\begin{equation} {\Theta} = {1/v \int {{G M }\over {r^2}} {{b}\over {r}} {{dx} \over {v}}}, \end{equation}

where b is the impact parameter and dx is an infinitesimal distance in the
velocity direction. In this simple case:

\begin{equation} {\Theta} = {{{1} \over {v^2}} {{2 G M} \over {b}}}. \end{equation}

The change in momentum in the direction of motion for this particle, $ \Delta p_{\parallel}$, is:

\begin{equation} \Delta p_{\parallel} = m v (cos \Theta  - 1), \end{equation}
which for small $\Theta$ can be expressed as:

\begin{equation} \Delta p_{\parallel} = - {{2 G^2 M^2 m } \over {b^2 v^3}}. \end{equation}

By conservation of momentum, $ \Delta p_{\parallel} = - \Delta P_{\parallel}$, 
\noindent so the large particle also experiences a backward deflection. The change
in velocity for this large mass is $ \Delta V = - \Delta p_{\parallel} / M $.

If the large mass were equally divided into n smaller masses such that 
the impact parameters were the same, equation A2 tells us that the 
deflection $\Theta$ would be:

\begin{equation} {\Theta} = {\sum\nolimits {{2 G M/n} \over {b v^2}}}, \end{equation}

\noindent where the sum is over the n particles. This deflection angle is equal to equation A2. Consequently, the momentum for any mass M is 
the same as it would be if the large mass were equally divided, despite the
apparent $M^2$ dependence. Likewise, if the small mass m were divided 
into n equal masses, the momentum of mass m as a whole is determined by 
the sum of n deflections, and is equivalent to the unpartitioned momentum. 

To get the force on the small mass, we can use Newton's 3rd law and 
find the acceleration of the large mass, $d V / dt$. When the small mass 
is subdivided into n particles with number density $\eta$, this acceleration is:

\begin{equation}  {{d V} \over {dt}} =  - \int {{{2 G^2 M m }\over {b^2 v^3}}\ 2 \pi b \ db \ \eta v }, \end {equation}

\noindent which simplifies to:

\begin{equation} {{d V} \over {dt}} = - {4 \pi G^2 M \rho \over {v^2}} \ ln \Lambda \end{equation}

\noindent where $ln \Lambda $ is the usual Coulomb logarithm.
For a gaussian spectrum of velocities, equation A6 results in the dynamical friction acceleration, $ dv_{f} / dt $, in 
equation 3 in the paper. Hence,
the dynamical friction force on a secondary can be equally apportioned
among equal mass secondary particles.

\newpage

\figcaption[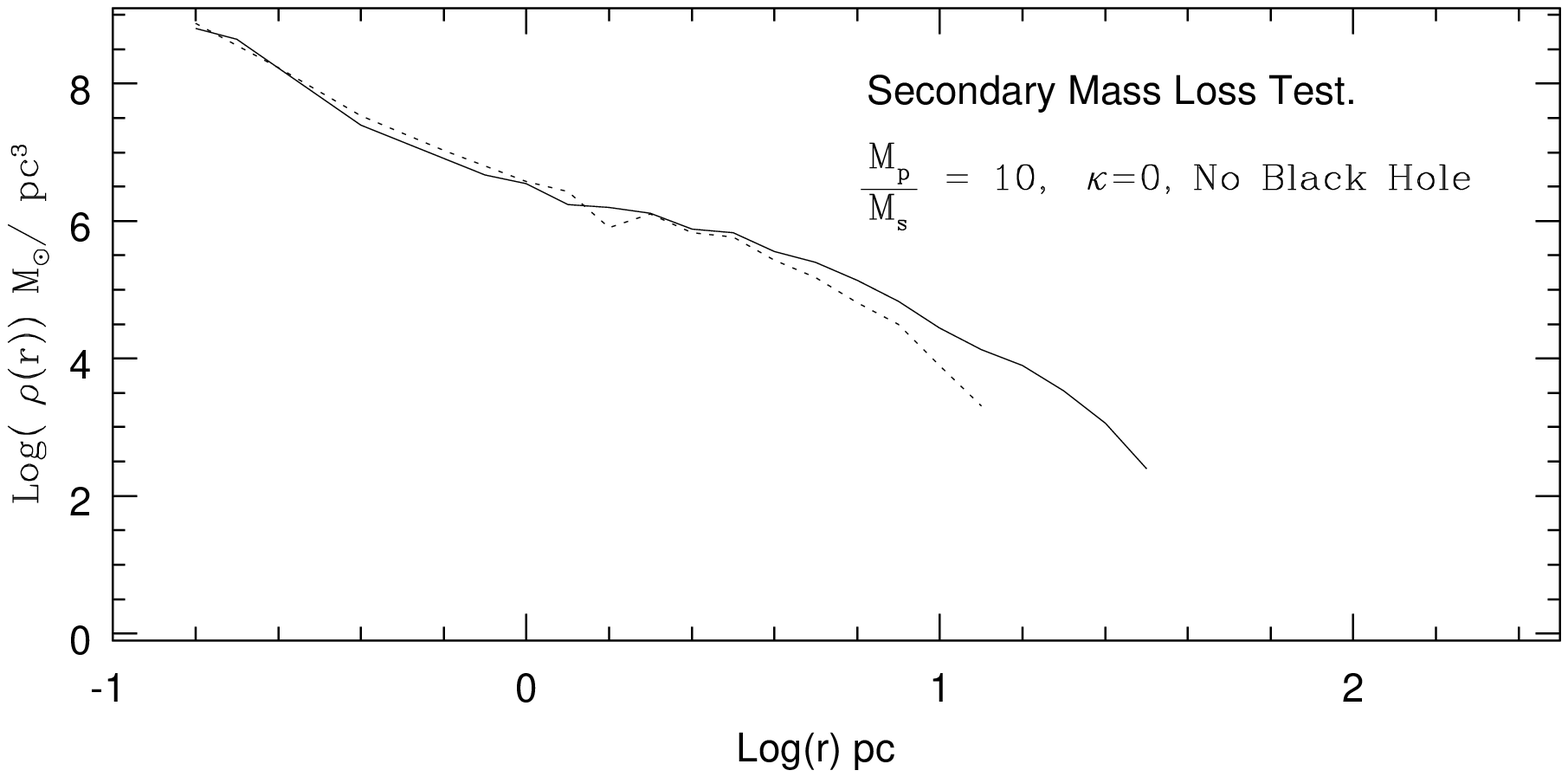]{Final secondary density for a purely stellar 10:1
merger with and without secondary mass loss. The density of the secondary 
is shown as a function of radius after the merger. The density 
which results from neglecting secondary mass loss in the orbital 
decay trajectory is a dashed line, and the density which results from
including this mass loss is a solid line. 
The changes are fluctuations due to small number statistics.
\label{Figure 1}}

\figcaption[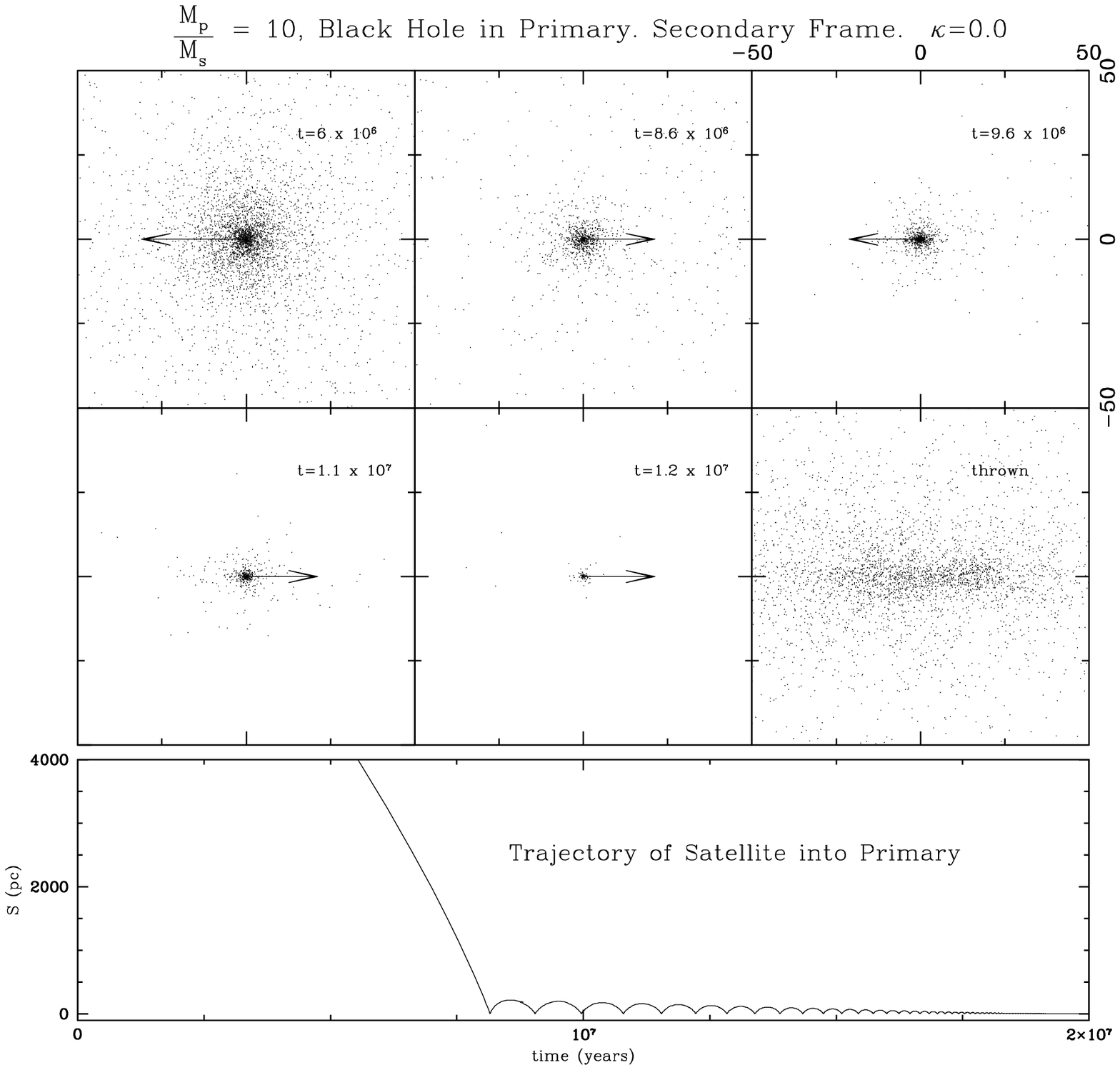]{The xy plane projection of a secondary as 
it merges with a primary 10
times more massive on a $\kappa=0.0$ encounter. Each panel represents 
a different snapshot of the bound 
secondary particles along its orbital decay trajectory. The last panel
represents the distribution of unbound particles by the end of the merger. 
The separation of the primary and secondary is shown on the bottom 
of the plot. Most of the envelope particles
are unbound after the first pass, and by the second pass, $90 \%$
of the matter is stripped away. However, the innermost particles
remain bound to the secondary's potential for several more passes, until the
secondary is entirely disrupted, and the debris is distributed over a 
large volume of space.
\label{Figure 2}}

\figcaption[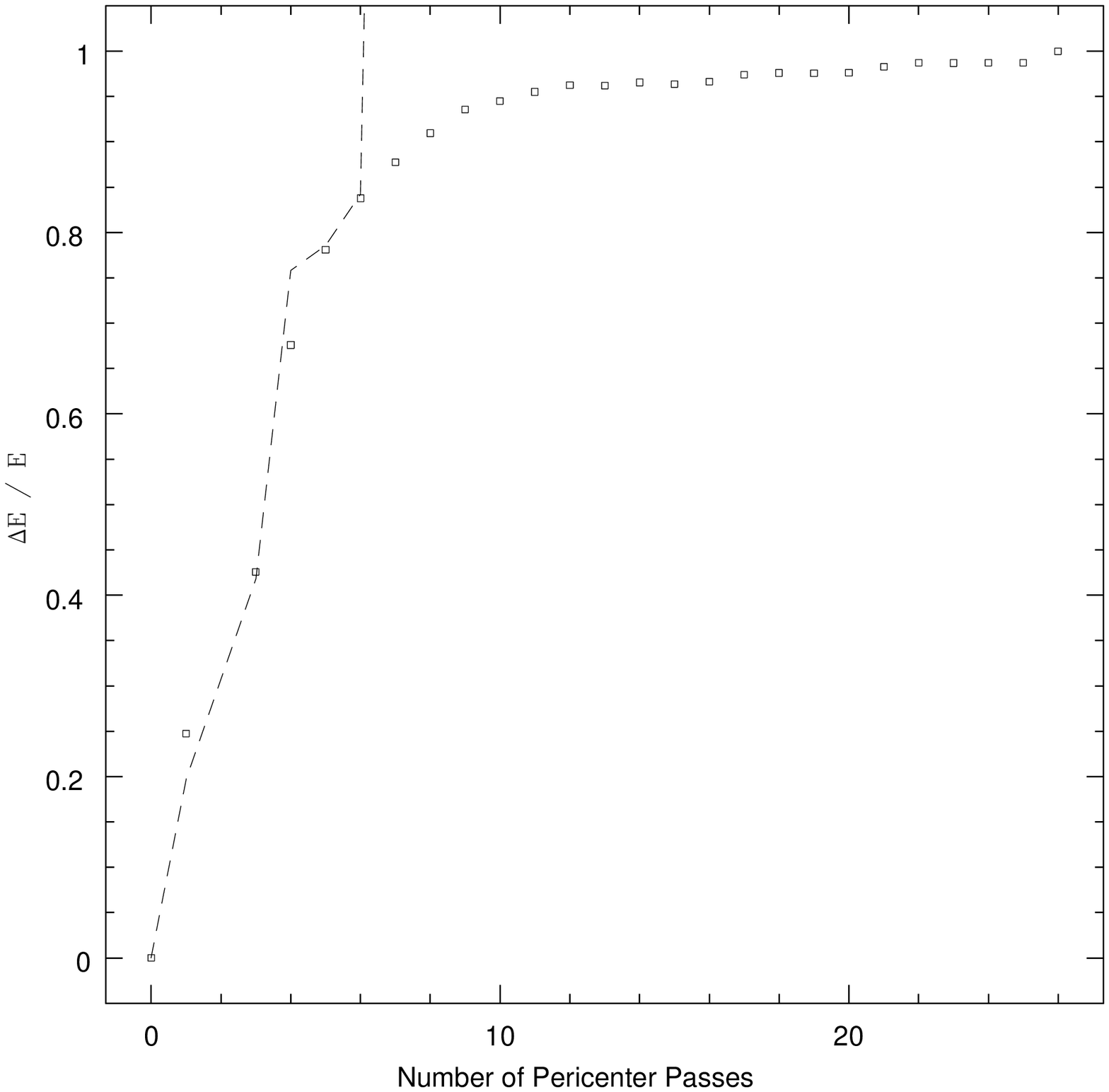]{The change in the secondary energy versus number 
of pericenter passes. We plot the change in the secondary energy 
(normalized by the initial secondary energy) versus the number of 
pericenter passes for the 10:1 $\kappa=0.0$ merger. The square points
are derived from our n-body simulation, and the solid line results from
an impulse approximation expression for the change in secondary energy 
induced by the black hole. The secondary is clearly disrupted after 
10  passes.
\label{Figure 3}}

\figcaption[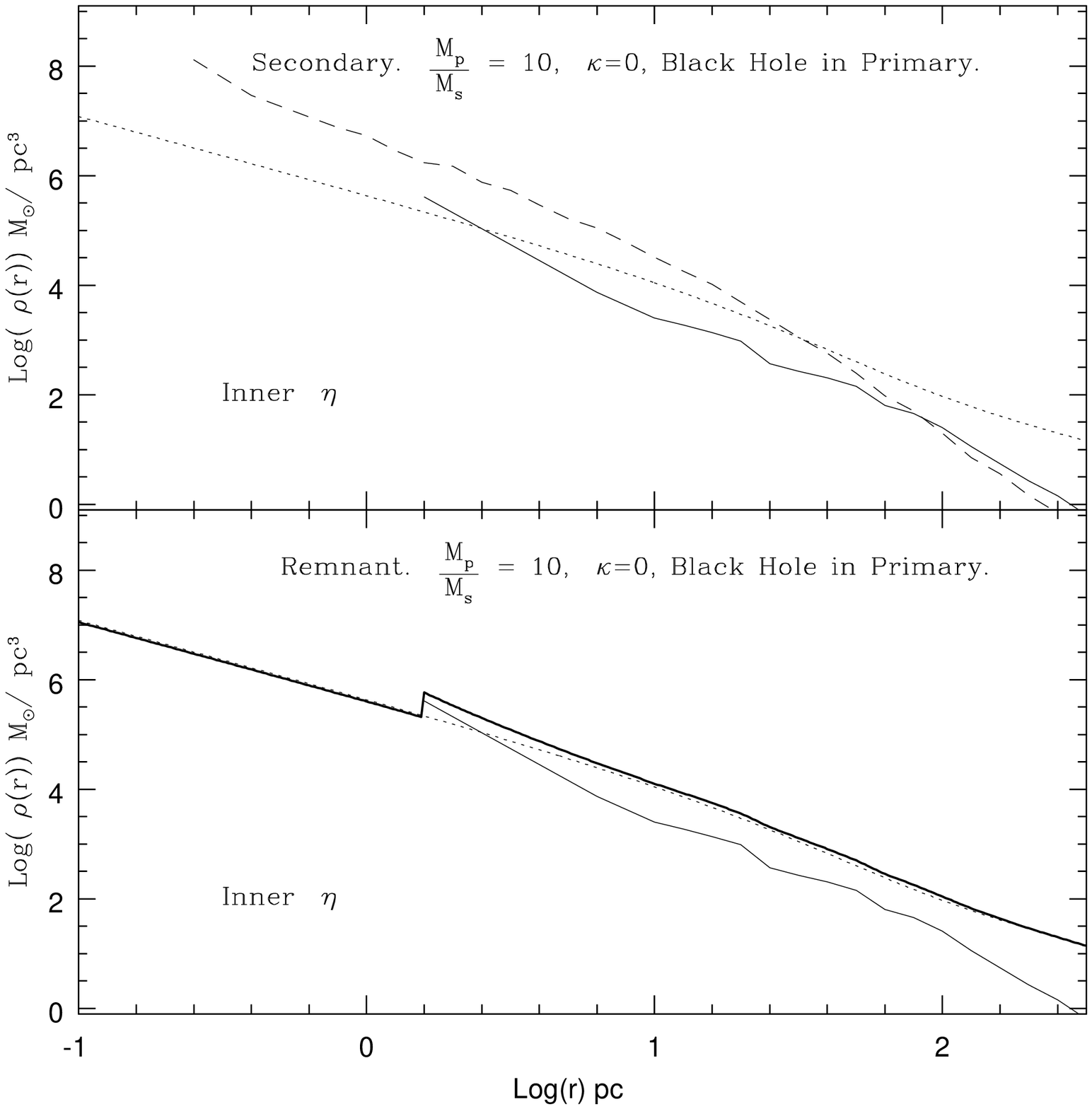]{Density profile for the mass ratio 10:1 
$\kappa=0.0$ encounter. In the top panel, we illustrate the change 
in the inner density profile of the secondary. We plot the density 
at a radius r against radius in parsecs. The solid line is the final 
secondary profile, the thick dashed line is the initial state of the 
secondary, and the dotted line is the density profile of the primary 
for comparison. In the bottom panel, the thick solid line represents 
the resulting remnant, the dotted line corresponds to the final state 
of the primary, and the thin solid line represents the final state of 
the secondary. The dramatically lower density in the secondary indicates 
it has been disrupted. \label{Figure 4}}

\figcaption[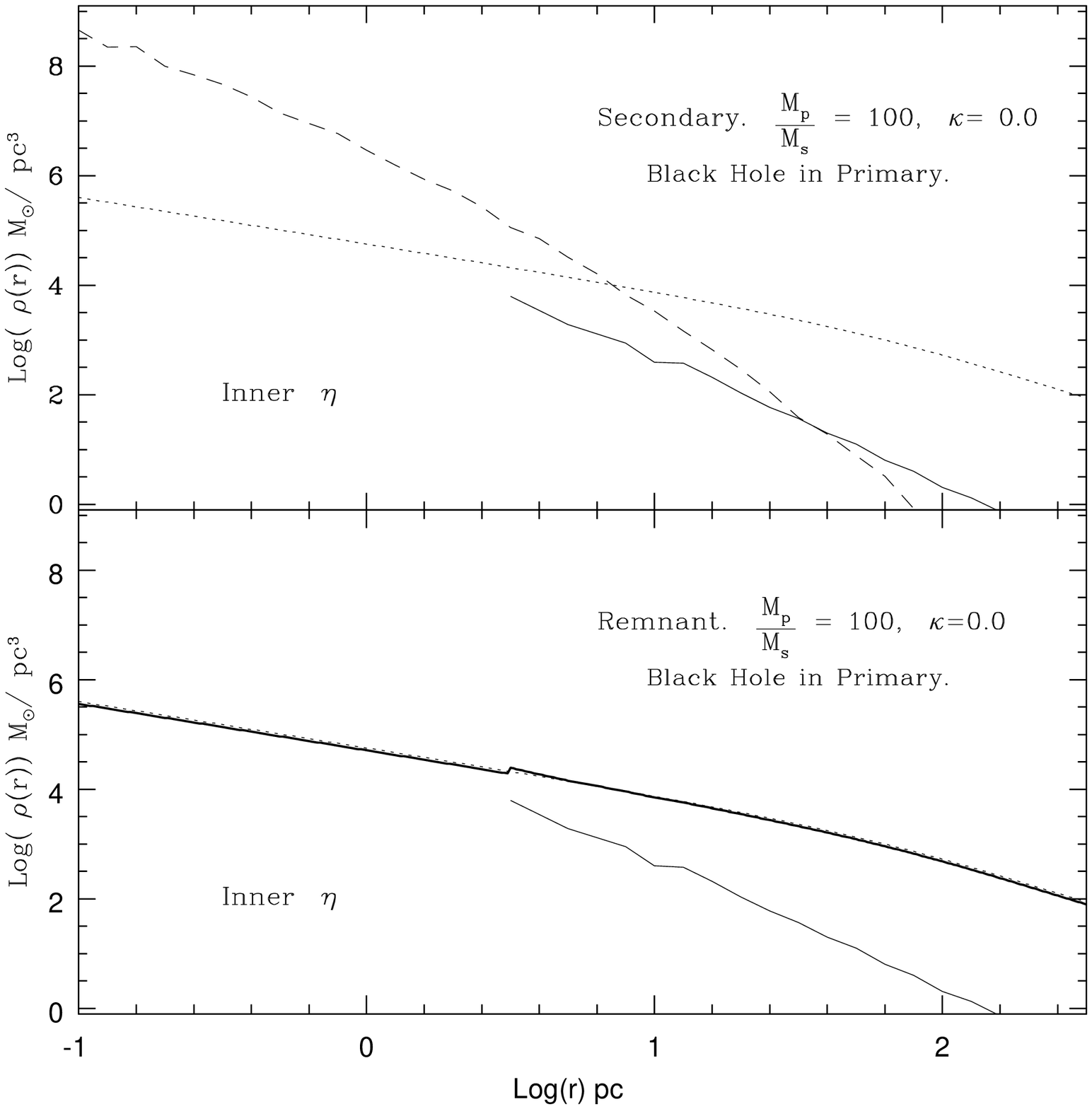]{Density profile for the mass ratio 100:1 
$\kappa=0.0$ encounter. See caption for figure 4. \label{Figure 5}} 

\figcaption[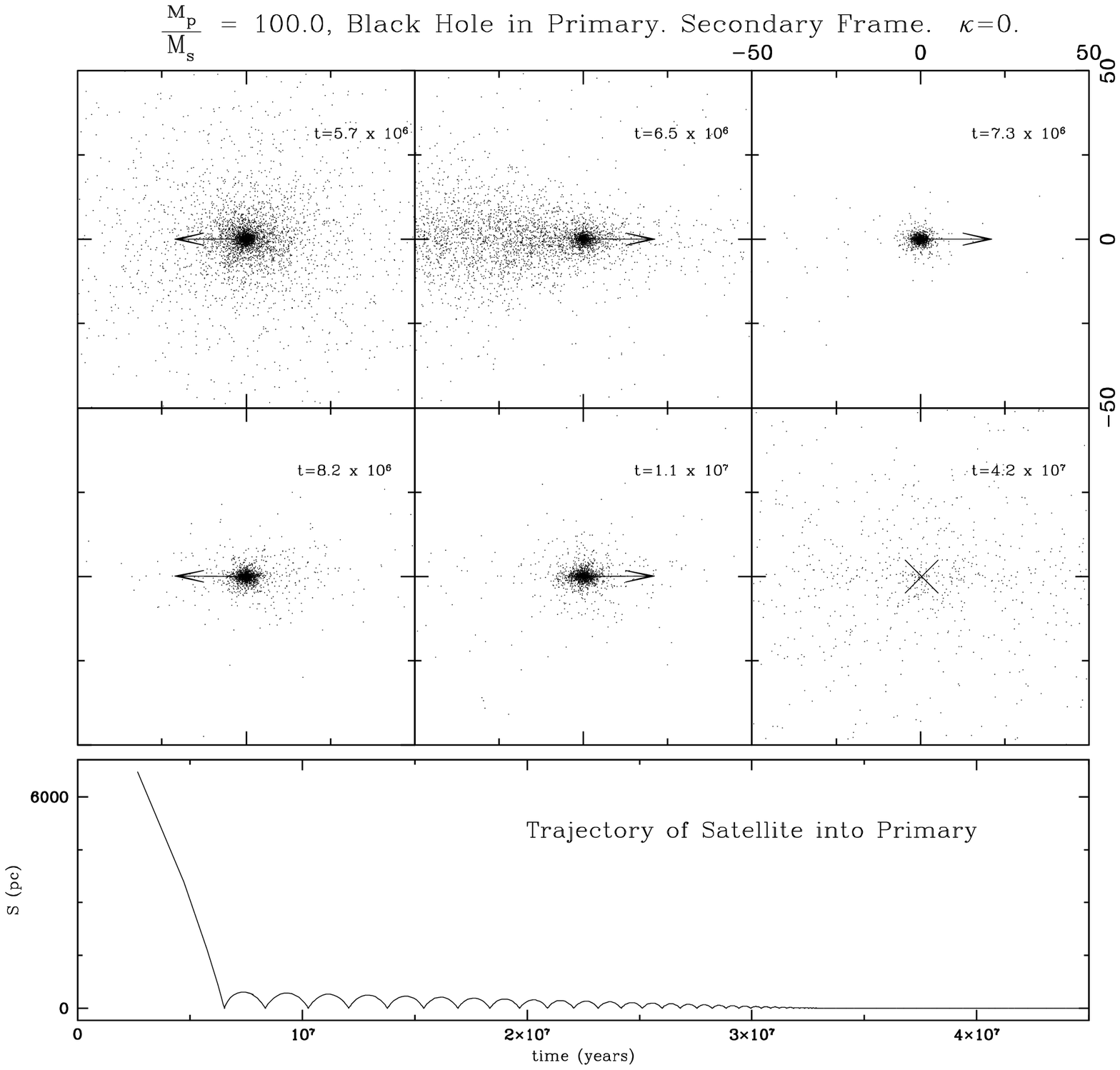] {The xy plane projection of a secondary as 
it merges with a primary 100 times more massive on a $\kappa=0.0$ 
encounter. See caption for figure 2.
\label{Figure 6}}

\figcaption[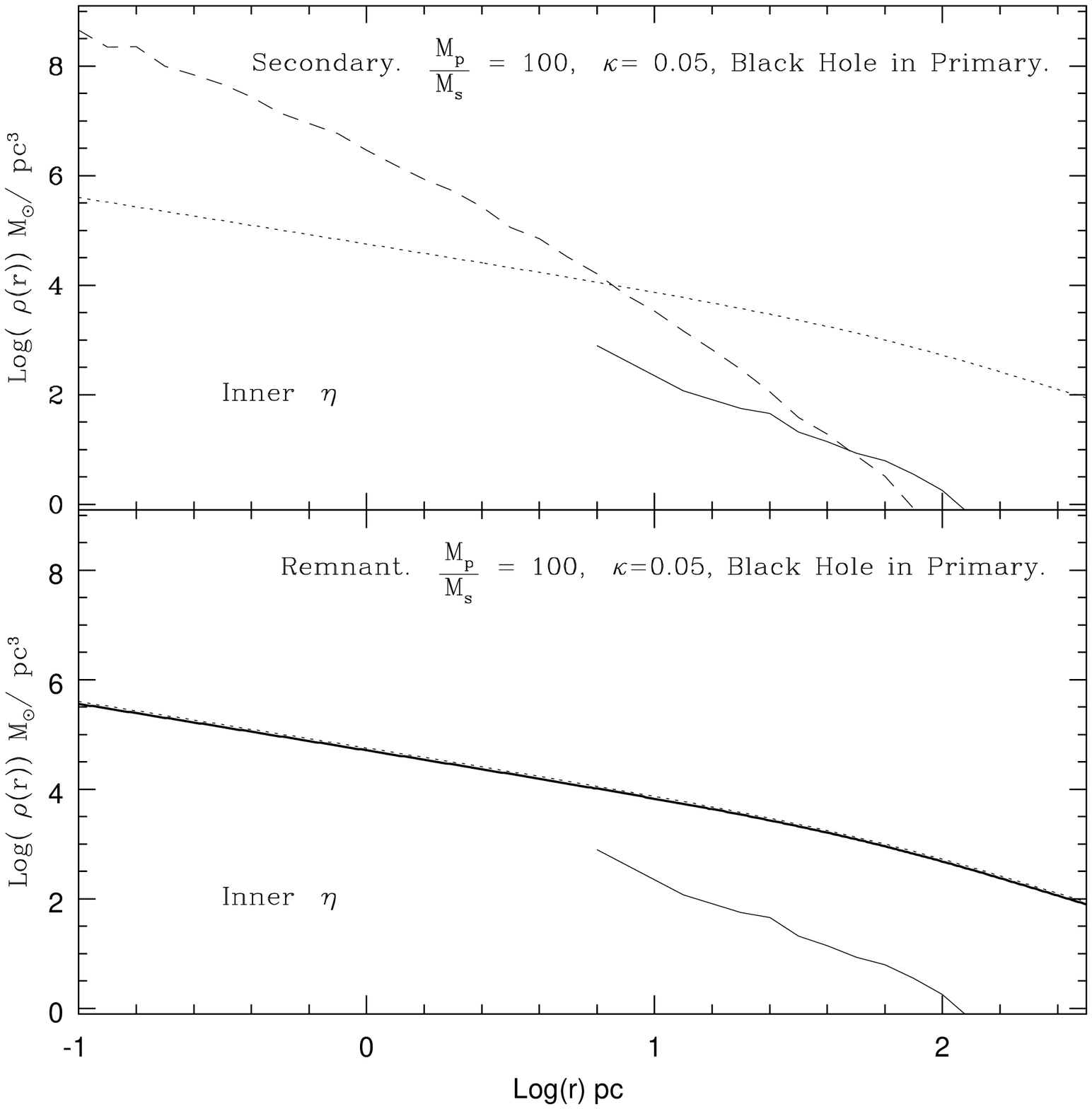] {Density profile for the mass ratio 100:1 $\kappa=0.05$
experiment. See caption for figure 4.
\label{Figure 7}}

\figcaption[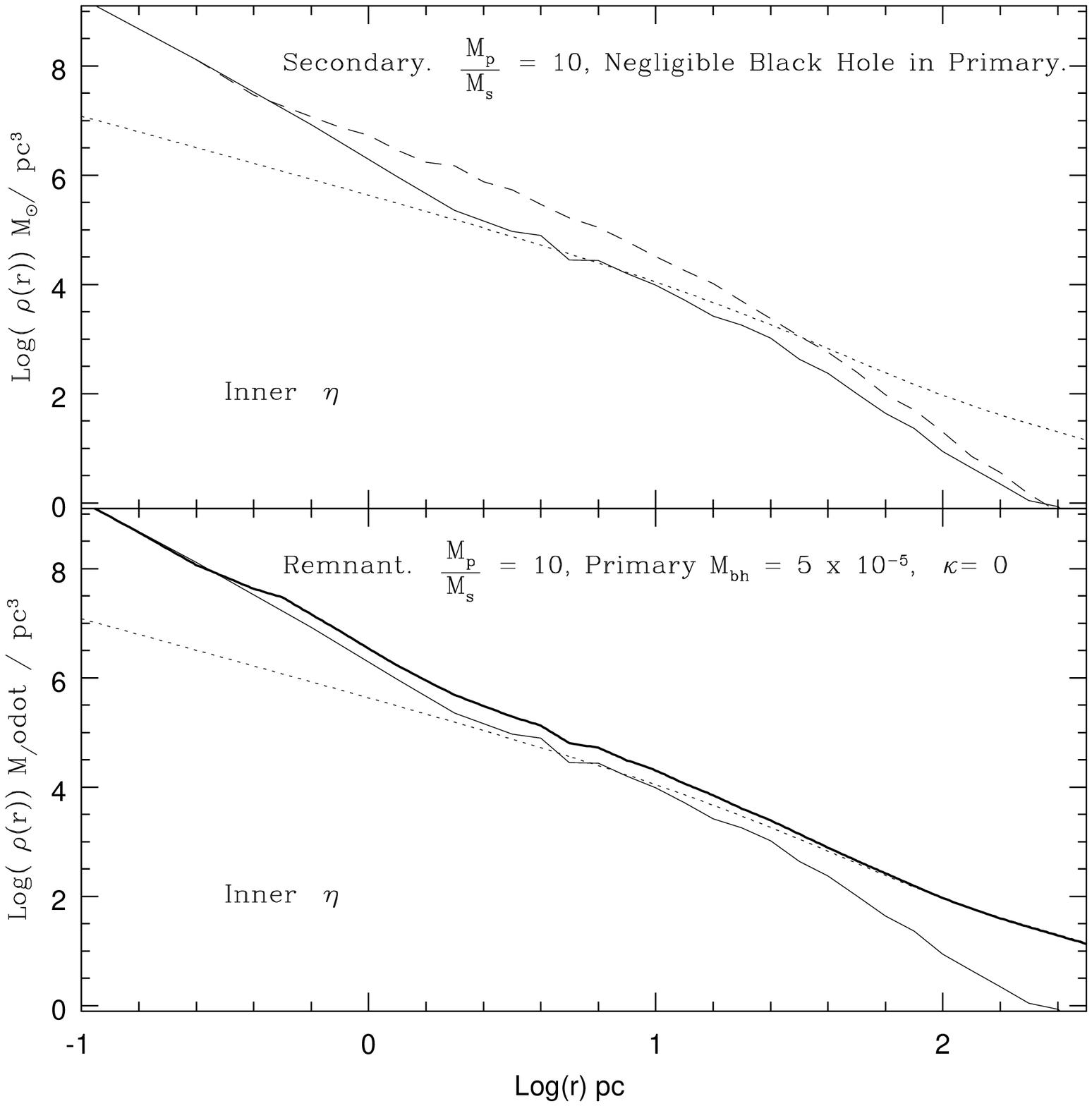]{Density profile for the mass ratio 10:1, $\kappa=0.0$
experiment with negligible primary black hole mass. See caption for
figure 4. Inside the tidal radius, the secondary is essentially intact.
\label{Figure 8}}

\figcaption[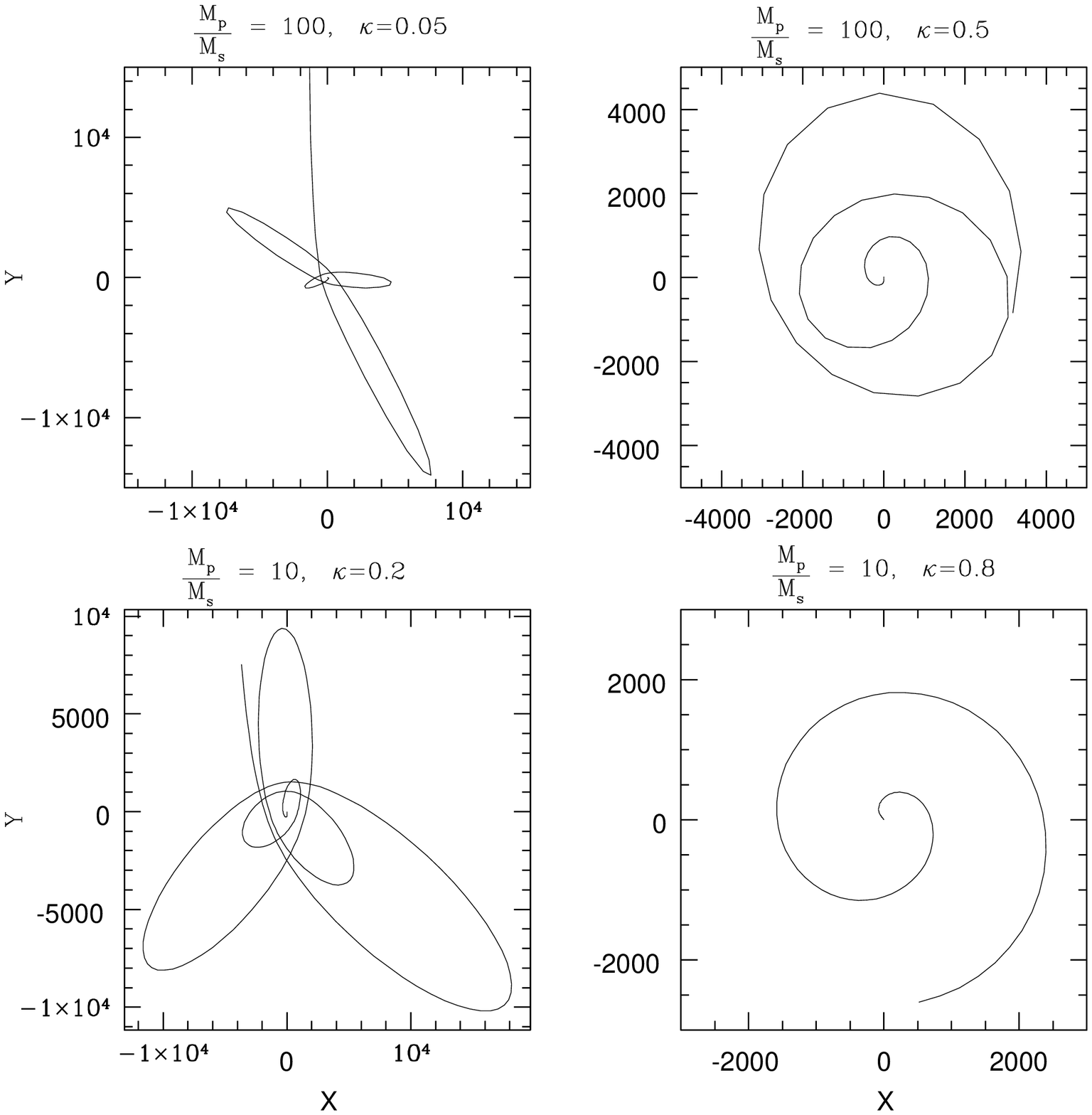]{The xy plane projection of orbital decay trajectories.
The orbital decay trajectories for the secondary in the 100:1 $\kappa=0.05$,
100:1 $\kappa=0.5$, 10:1 $\kappa=0.2$, and 10:1 $\kappa=0.8$ experiments are
shown in panels A, B, C, and D, respectively.
\label{Figure 9}}

\figcaption[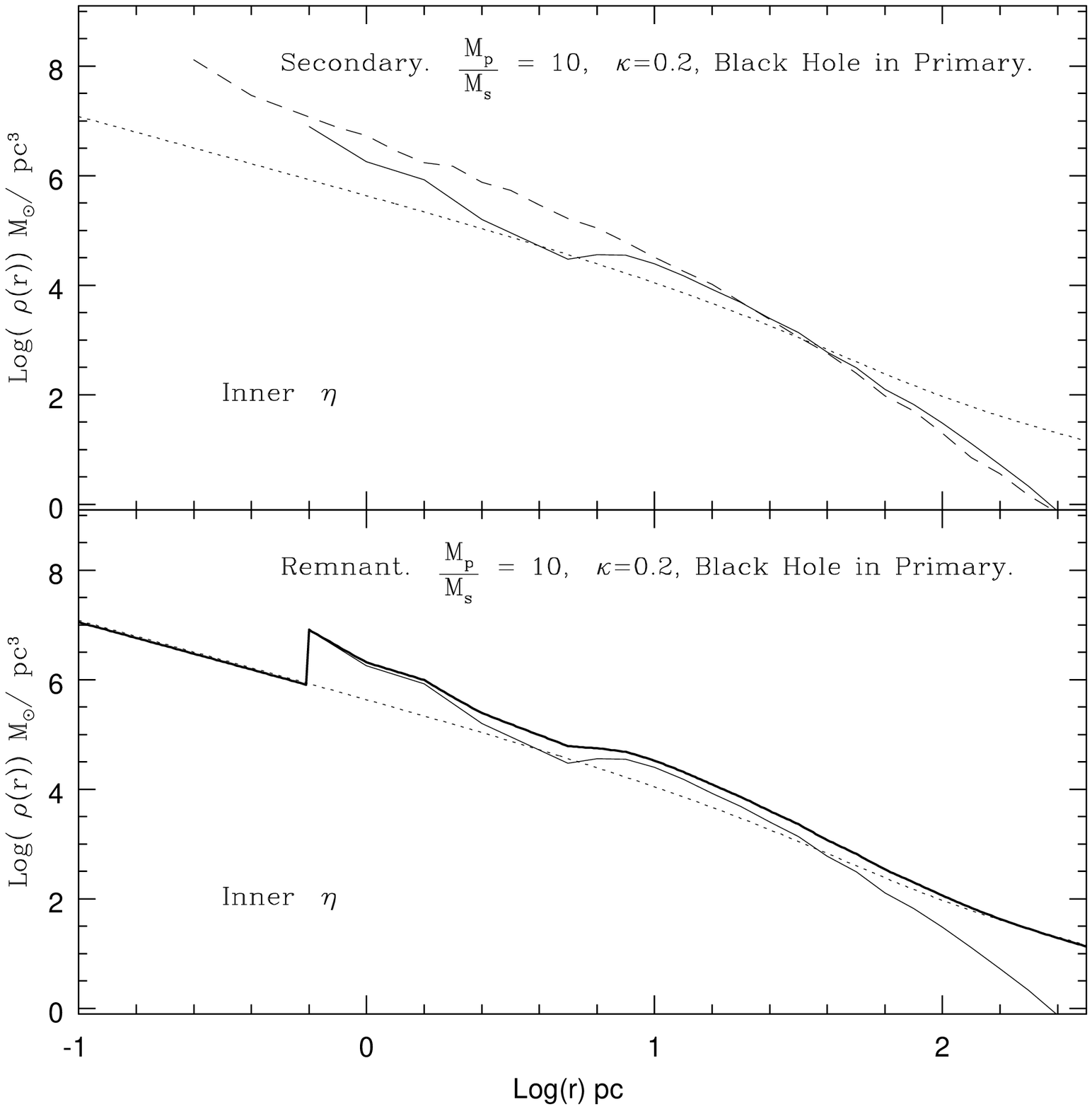]{Density profile for the 10:1, $\kappa=0.2$ experiment
See caption for figure 4.
\label{Figure 10}}

\figcaption[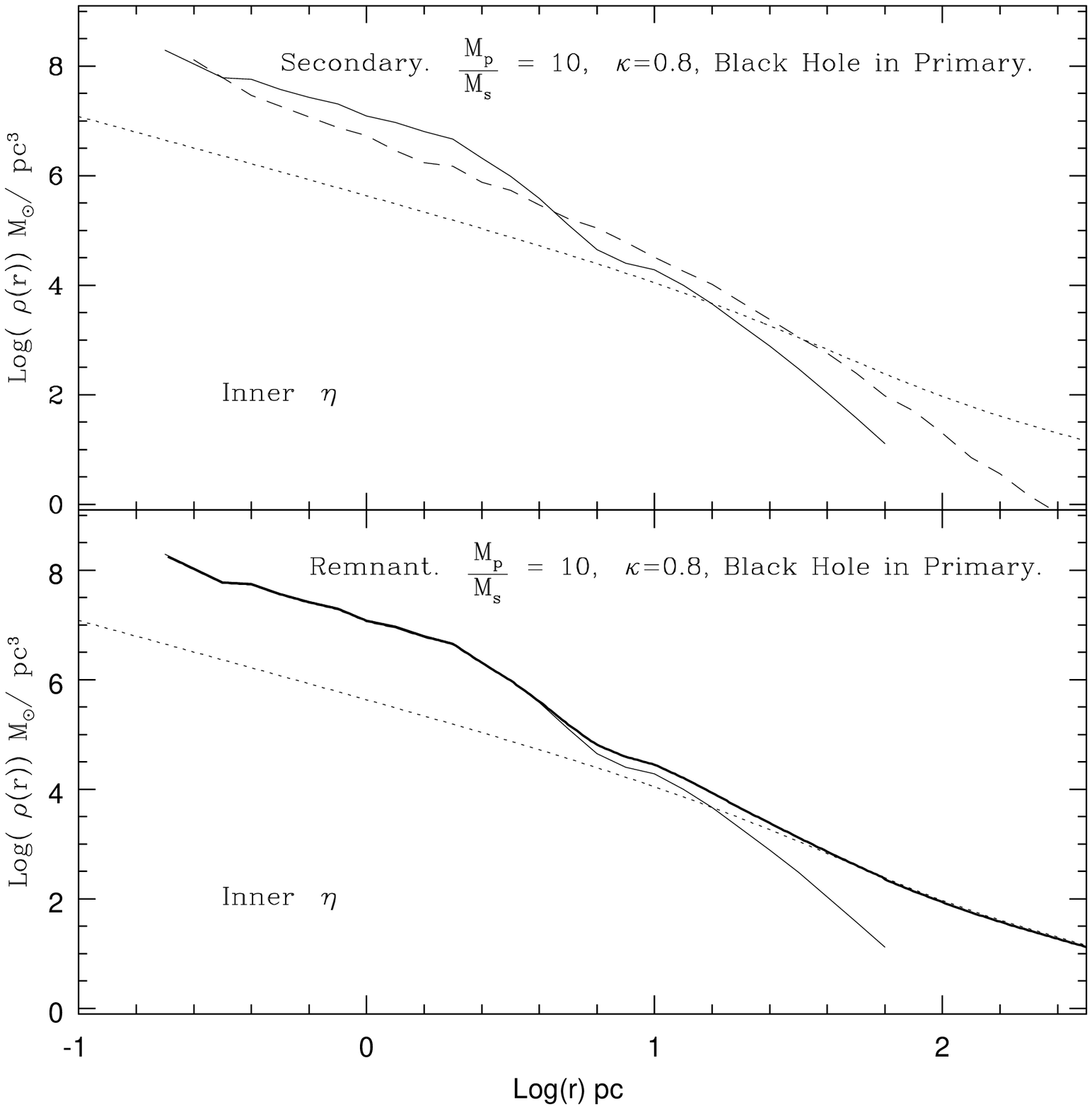]{Density profile for the 10:1, $\kappa=0.8$ experiment
See caption for figure 4.
\label{Figure 11}}

\figcaption[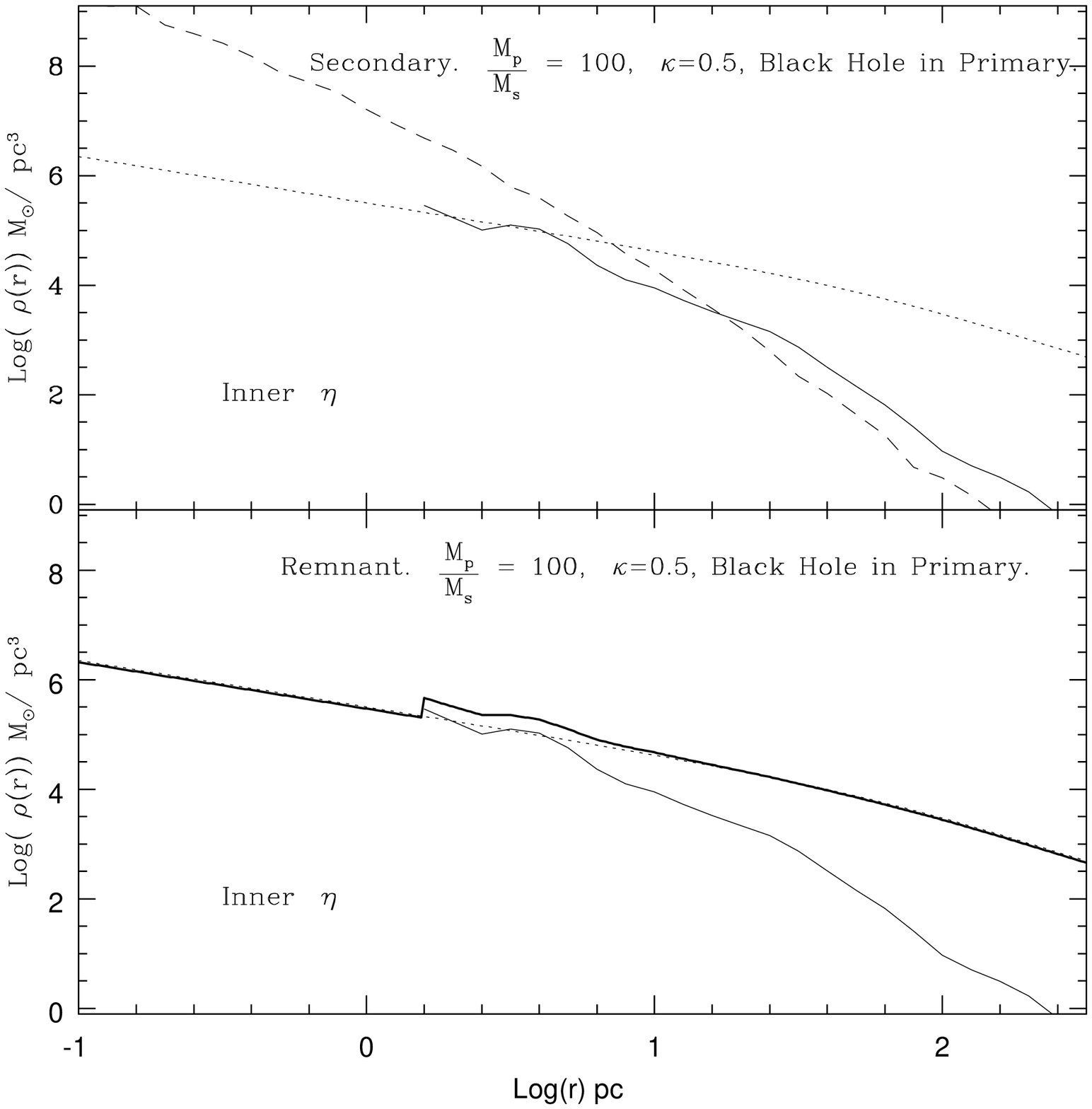]{Density profile for the 100:1, $\kappa=0.5$ encounter.
See caption for figure 4.
\label{Figure 12}}

\figcaption[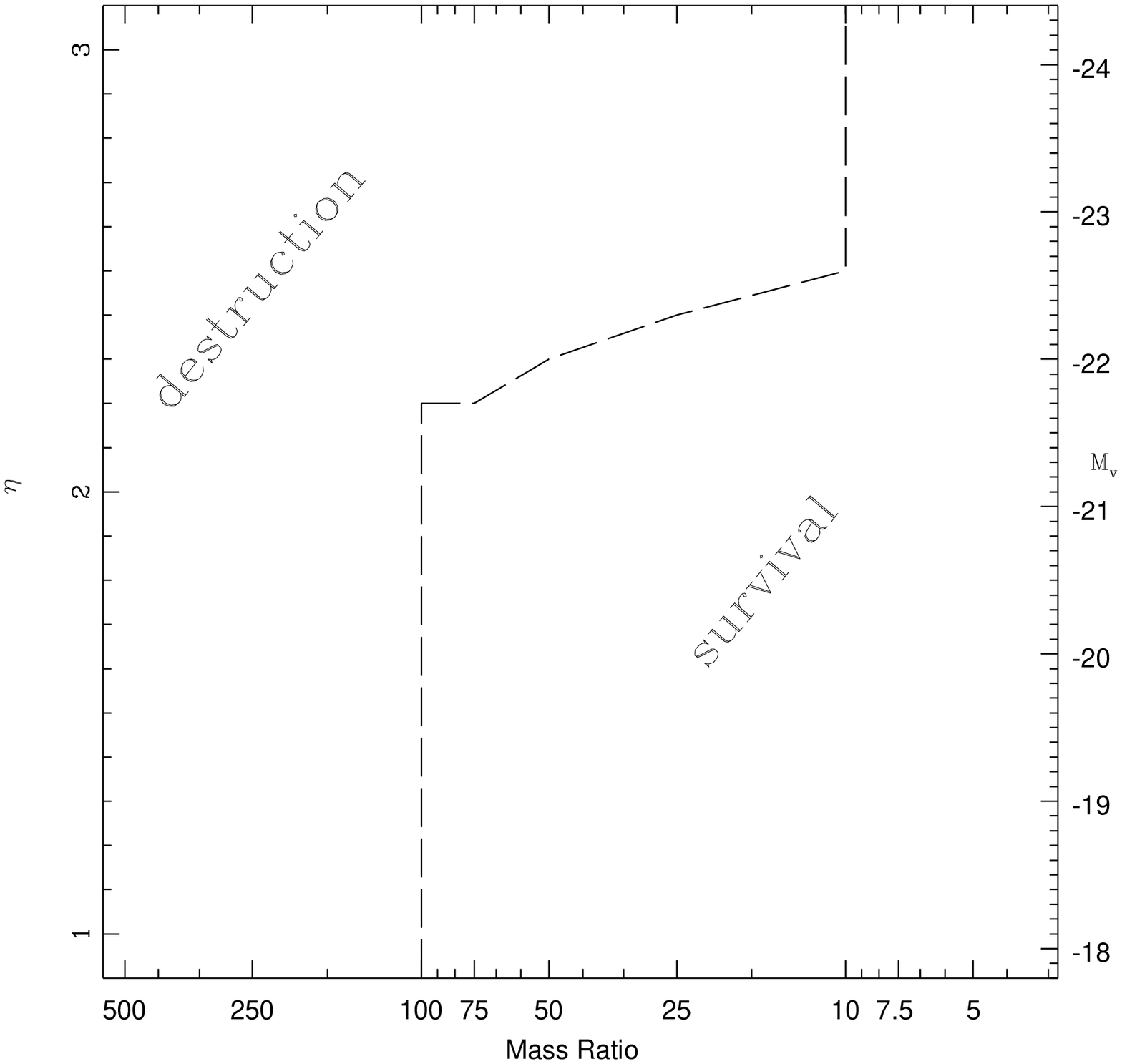]{Prediction of final secondary state for the
$\kappa=0.5$ orbit as a function of primary galaxy mass and central 
density slope. The dashed line marks the boundary between secondary
destruction and survival.
\label{Figure 13}}

\figcaption[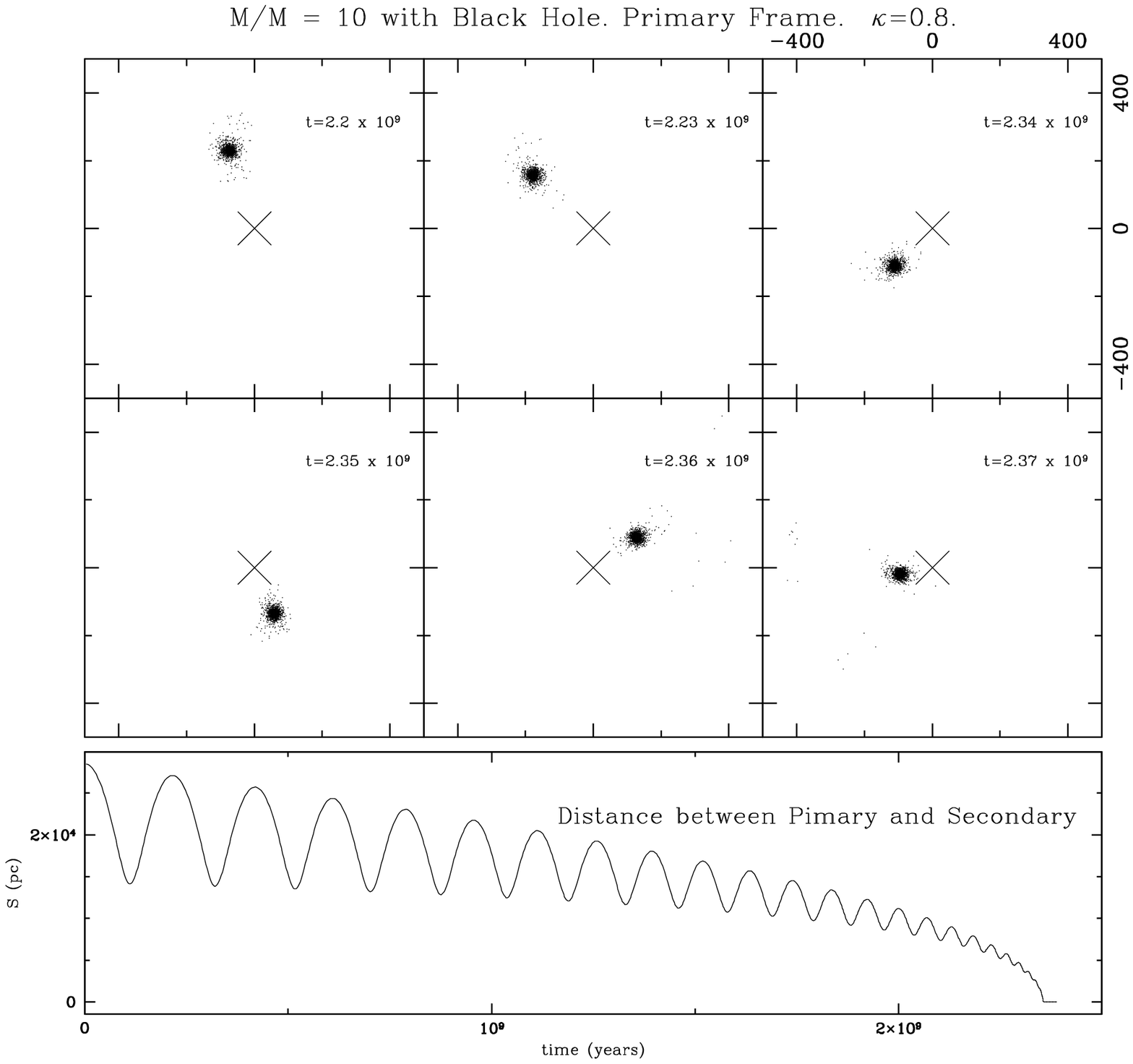]{The xy projection of the secondary galaxy viewed in
the primary frame for the 10:1, $\kappa=0.8$ experiment. See caption for
figure 2. The leading direction of the secondary distortion indicates that the
secondary is spinning, and that the spin may be induced by the black hole.
\label {Figure 14}}

\figcaption[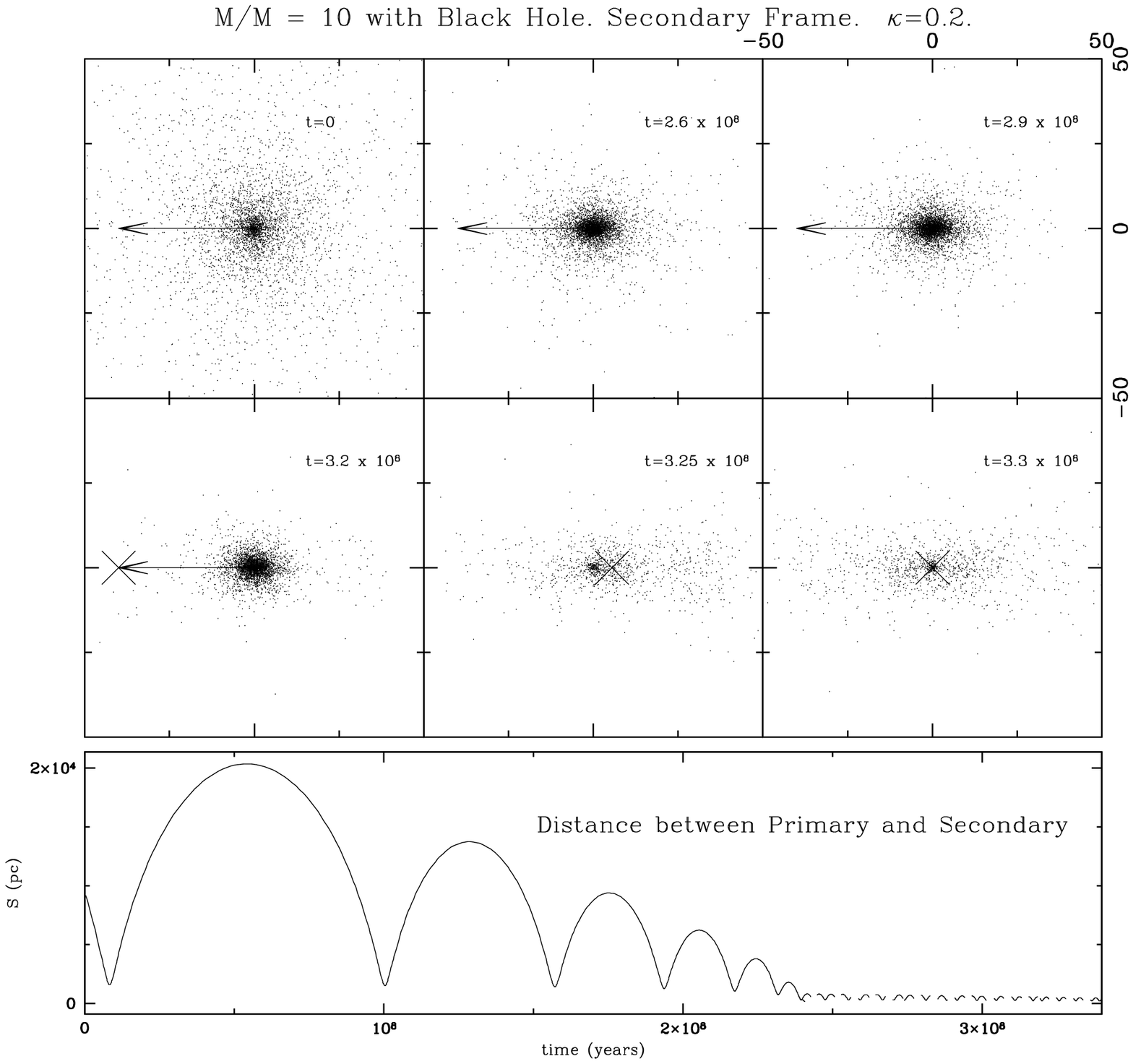]{The xz plane projection of a secondary as it merges
with a primary 10 times more massive on a $\kappa=0.2$ orbit. See caption
for figure 2. The secondary experiences significant flattening during the
final pericenter pass (frame 4), and the particles are unbound to 
to secondary once it has settled to the primary center (frame 5). The last
frame illustrates that the debris remains tightly bound to the primary
center.
\label{Figure 15}}

\figcaption[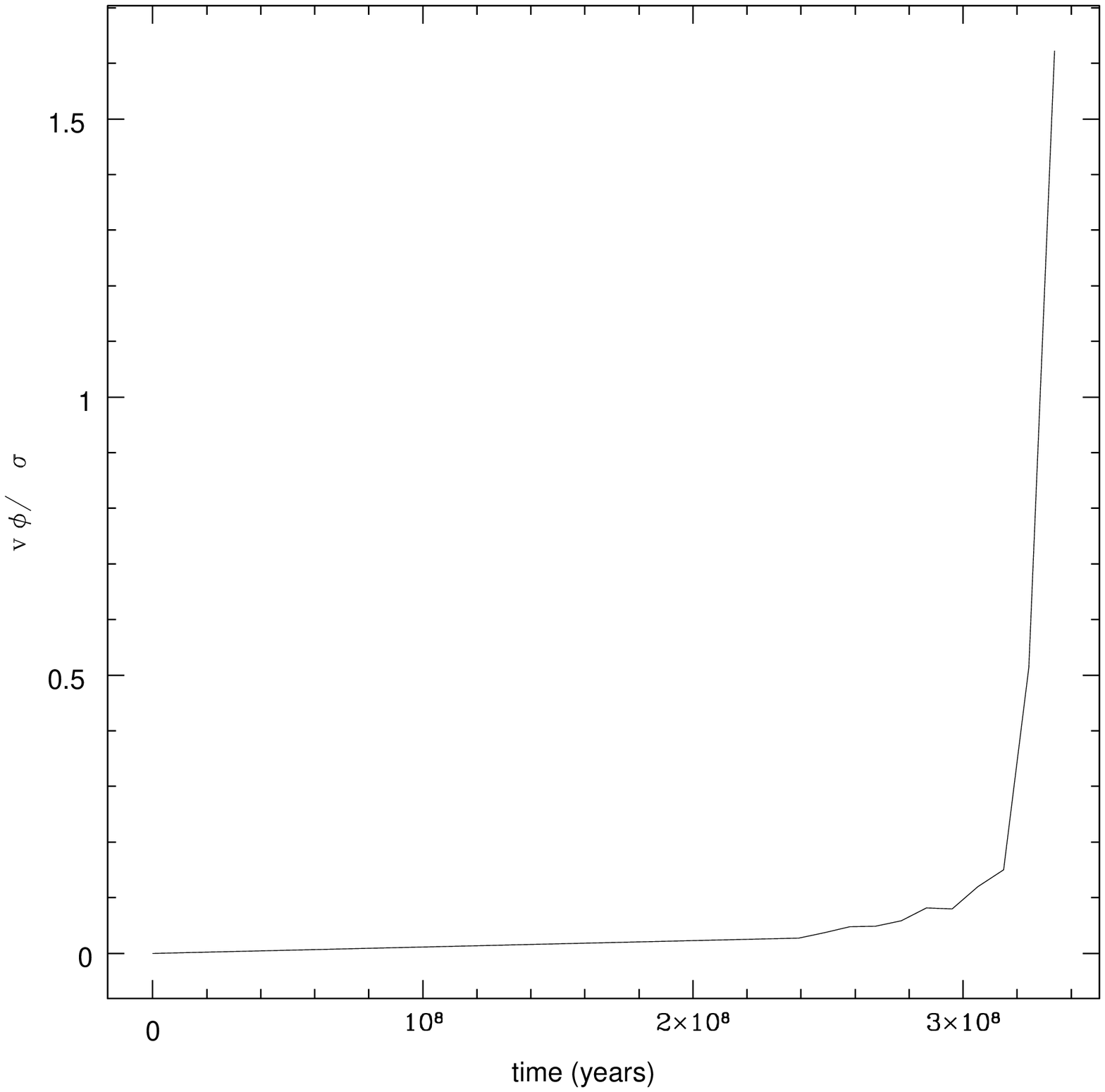]{The increase in secondary spin for the 10:1, $\kappa=0.2$
experiment. We plot the azimuthal velocity of the secondary over the
velocity dispersion versus time for a portion of the merger. The rapid
increase of the circular velocity indicates that the secondary is
spinning up over this time period.
\label{Figure 16}}

\figcaption[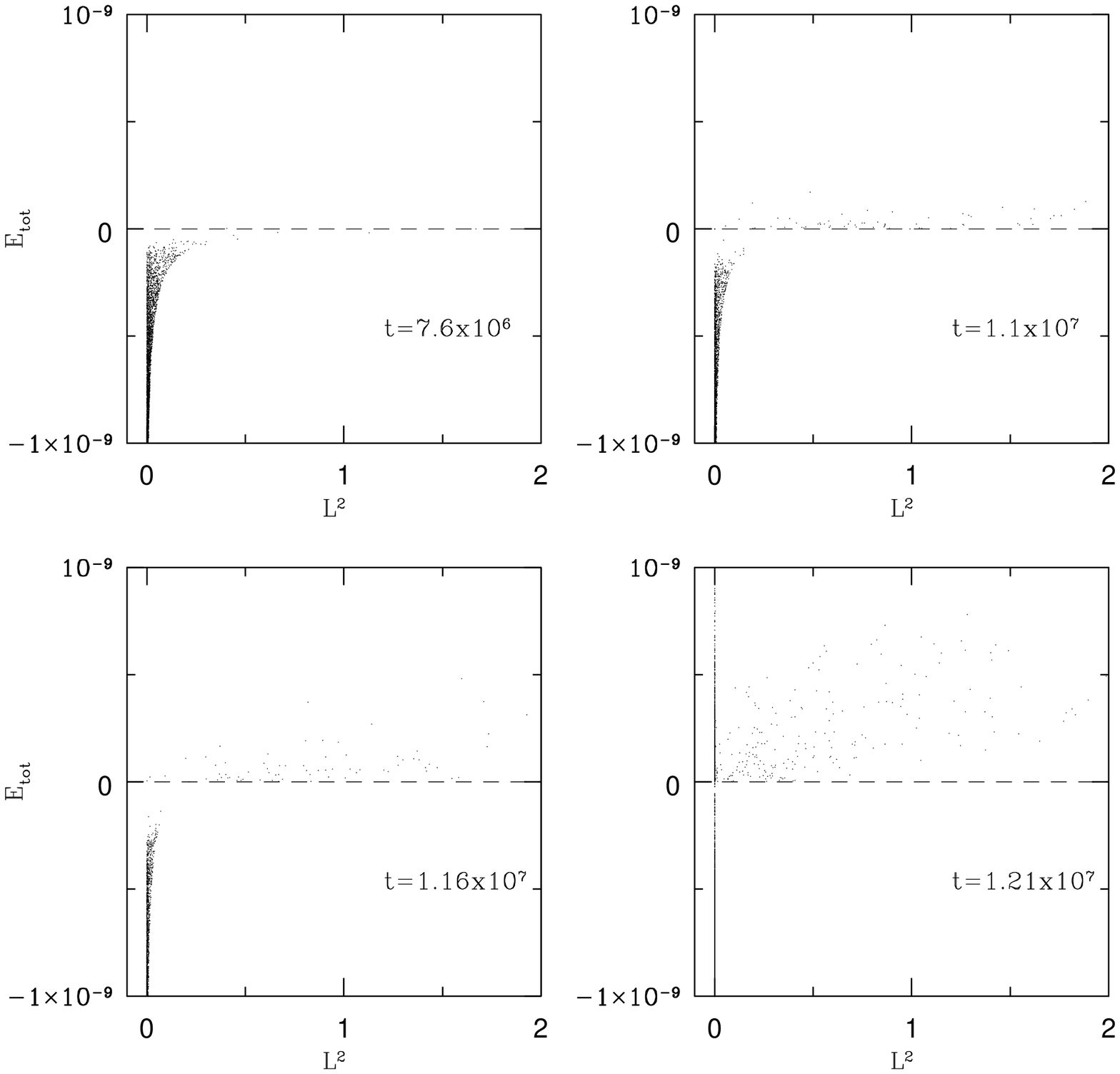]{The preferential loss of high angular momentum 
particles. We plot the energy versus the angular momentum of secondary
particles in the secondary frame for different times in the 10:1, $\kappa=0.2$
experiment. The angular momentum of the secondary sharply increases during
the last orbit, and the particles with the highest angular momentum
become unbound. In the last frame, the high angular momentum particles
were stripped before the secondary reached the primary center. A few 
secondary crossing times after the secondary reached the primary center, the
more plunging orbits were stripped, but still remained tightly bound to the
primary.
\label{Figure 17}}

\newpage



\makeatletter
\def\jnl@aj{AJ}
\ifx\revtex@jnl\jnl@aj\let\tablebreak=\nl\fi
\makeatother


\begin{deluxetable}{lrrrrcrrrrr}
\scriptsize
\tablewidth{0pc}
\tablecaption{Eta Model Galaxy Parameters}
\tablehead{
\colhead{${M_1}\over{M_2}$}     
& \colhead{Galaxy}      
&\colhead{M$_{\rm v}$}          
& \colhead{r$_{\rm core}$\tablenotemark{a} }       
&\colhead{M$_{\rm core}$\tablenotemark{b} }     
& \colhead{$\eta_{ \rm core}$ }    
&\colhead{r$_{\rm env}$ }      
& \colhead{M$_{\rm env}$ }        
&\colhead{$\eta _{\rm env}$ }   
& \colhead{r$_{\rm half}$ }    
&\colhead{M$_{\rm bh}$} }
\startdata
100:1 & Primary& -22.0& 263& 4.7 x 10$^{10}$& 2.15 & 4000 &  3.4 x 10$^{12}$ & 2.2 & 10617 & 6.3 x 10$^{9}$ \nl
&&&&1.34&&&98.66&&&0.18\nl
& Secondary& -18.0& 3.8& 1.6 x 10$^{8}$ & 1.0 & 300 & 3.5 x 10$^{10}$ & 1.5 & 508 &\nl
&&&&4.4 x 10$^{-3}$&&&0.996&&\nl
10:1 & Primary& -21.5& 155& 2.3 x 10$^{10}$ & 1.96 & 4000 & 1.9 x 10$^{12}$ &2.0 & 9506 & 3.2 x 10$^{9}$\nl
&&&&0.117&&&9.88&&&0.02\nl
& Secondary& -19.5 & 18.63 & 1.3 x 10$^{9}$ & 1.23 & 1029 & 1.95 x 10$^{11}$ & 1.5& 1734 &  \nl
&&&&6.7 x 10$^{-3}$ &&& 0.993 &&\nl
 
\tablenotetext{a}{radii are in units of pc.}
\tablenotetext{b}{the top masses are in units of M$_{\odot}$, and the bottom masses are normalized such that the total secondary mass is 1.0}
\enddata
\end{deluxetable}

\begin{deluxetable}{lrcr}
\scriptsize
\tablewidth{0pc}
\tablecaption{Secondary Spatial Resolution}
\tablehead{
\colhead{${M_1}:{M_2}$ }
& \colhead{Type}
& \colhead{Particle Number}
& \colhead{Resolution (pc)}}
\startdata
100:1   & Double $\eta$ & 5000 & 10.08 \nl
	& Inner $\eta$ &  2000 & 0.07 \nl
	& 	       & 5000 & 0.05  \nl
\nl
10:1 	& Double $\eta$ & 5000 & 37.79 \nl
	& Inner $\eta$   & 2000 & 0.57 \nl
        & Inner $\eta$  & 5000 & 0.42 \nl
\nl
2.5:1	& Double $\eta$ & 5000 & 119.6 \nl
	& Inner $\eta$  & 2000 & 3.04 \nl
        & Inner $\eta$  & 5000 & 2.24 \nl
\enddata
\end{deluxetable}

\begin{deluxetable}{lrcrr}
\scriptsize
\tablewidth{0pc}
\tablecaption{Synopsis of Experiments}
\tablehead{
\colhead{Mass Ratio}
& \colhead{Initial apo:peri}
& \colhead{$\kappa$}
& \colhead{Black Hole Mass}
& \colhead{Effect on Secondary}}
\startdata
100:1 & 31850:0 & 0.0 & 0.18 & Disrupted \nl
& 31850:460 & 0.05 & 0.18 & Disrupted \nl
10:1 & 28520:0 & 0.0 & 0.02 & Disrupted \nl
&  28520:0 & 0.0 & 0.0005 & Intact \nl
100:1 & 31850:8650 & 0.5 & 0.18 & Disrupted \nl
10:1& 28520:1640 & 0.2 & 0.02 & Intact, spinning \nl
& 28520:14280& 0.8 & 0.02 & Intact, spinning \nl
\enddata
\end{deluxetable}

\end{document}

%% file: abbrevs
 \def\today{\number\day\enspace
      \ifcase\month\or January\or February\or March\or April\or May\or
      June\or July\or August\or September\or October\or
      November\or December\fi \enspace\number\year}
 \def\clock{\count0=\time \divide\count0 by 60
     \count1=\count0 \multiply\count1 by -60 \advance\count1 by \time
     \number\count0:\ifnum\count1<10{0\number\count1}\else\number\count1\fi}
 
 \def\newline{\hfil\break}

\def\et{{\it et~al. }}

 \def\bul{\ifmmode\bullet\else$\bullet$\fi}

 \def\deg{\ifmmode^\circ\else$^\circ$\fi}

 \def\kms{\ifmmode \hbox{ \rm km s}^{-1} \else{ km s$^{-1} $}\fi} 

\def\sec{\ifmmode \hbox{\rm sec}\else{sec}\fi} 
\def\yr{\ifmmode \hbox{\rm yr}\else{yr}\fi} 
\def\myr{\ifmmode \hbox{\rm Myr}\else{Myr}\fi} 
\def\gyr{\ifmmode \hbox{\rm Gyr}\else{Gyr}\fi} 

\def\hz{\ifmmode \hbox{\rm hz}\else{hz}\fi} 
 
\def\kpc{\ifmmode \hbox{\rm  kpc}\else{kpc}\fi} 
 \def\pc{\ifmmode \hbox{\rm  pc} \else{pc}\fi} 
\def\mpc{\ifmmode \hbox{\rm Mpc} \else{Mpc}\fi} 

\def\erg{\ifmmode \hbox{\rm erg} \else{erg}\fi} 

 \def\yrs{\ifmmode \hbox{\rm yrs}\else{yrs}\fi}

 \def\angstr{\ifmmode{\rm \AA}\else\AA\fi}

\def\ho{\ifmmode H_0\else$H_0$\fi}
\def\omo{\ifmmode\Omega_0\else$\Omega_0$\fi}
\def\to{\ifmmode T_0\else$T_0$\fi}
\def\h-1{h^{-1}}

\def\d2#1#2{{d^2#1 \over d#2^2}}

%% file: ms.bbl
\begin{references}
\reference{} Aarseth, S. J., \& Fall, S. M., 1980, ApJ, 236, 43
\reference{} Ashman, K., \& Zepf, S., 1992, ApJ, 384, 50
\reference{} Capaccioli, M., Held, E., Nieto, J.-L. 1987, AJ, 94, 1
\reference{} Carollo, C., Franx, M., Illingworth, G. D., Forbes, D. 1997, ApJ, 481, 710
\reference{} Faber, S., Kormendy, J., Byun, Youg-Ik, Dressler, A., Grillmair, C., Lauer, T., Richstone, D., Gebhardt, K., Tremaine, S. 1997, AJ, 114, 1771
\reference{} Ferrarese, L., Ford, H. C., Jaffe, W. 1996, ApJ, 470, 444
\reference{} Forbes, D. A., \& Thomson, R. C. 1992, MNRAS, 254, 723
\reference{} Franx, M., \& Illingworth, G. D. 1988, ApJ, 327, 55
\reference{} Governato, F., Colpi, M., Maraschi, L. 1994, MNRAS, 271, 317
\reference{} Holley-Bockelmann, K., \& Richstone, D. 1999, ApJ, 517, 92
\reference{} Kissler-Patig, M., Forbes, D., Minniti, D. 1998, MNRAS, 298, 1123
\reference{} Kormendy, J. 1984, ApJ, 287, 577
\reference{} Kormendy, J. 1985, ApJL, 292, L9
\reference{} Kormendy, J. \& Richstone D. 1995, Ann Rev Astron and Astroph, 33, 581 
\reference{} Kormendy, J., Bender, R., Richstone, D., Ajhar, E. A., Dressler, A., Faber, S. M., Gebhardt, K., Grillmair, C., Lauer, T., Tremaine, S. 1996, ApJ, 459, L57
\reference{} Lauer, T. 1985, ApJ, 292, 104 
\reference{} Lauer, T., Tremaine, S., Ajhar, E., Bender, R., Dressler, A., Faber, S., Gebhardt, K., Grillmair, C., Kormendy, J., Richstone, D. 1996, ApJ, 471, 79
\reference{} Makino, J., \& Ebisuzaki, T. 1996, ApJ, 465, 527
\reference{} Malin, D. F., \& Carter, D. 1983, ApJ, 274, 534
\reference{} Merritt, D., \& Quinlan, G. 1998, ApJ, 498, 625
\reference{} Merritt, D., \& Valluri, M. 1998, ApJ, 509, 933
\reference{} Quinlan, G., \& Hernquist, L. 1997, New Astronomy, 2, 533
\reference{} Richstone D., Ajhar, E. A., Bender, R., Bower, G., Dressler, A., Faber, S., Filippenko, A. V., Gebhardt, K., Green, R., Ho, L. C., Kormendy, J., Lauer, T. R., Magorrian, J., Tremaine, S. 1998, Nature, 1998, 395, 14
\reference{} Schweizer, F. 1982, ApJ, 252, 455
\reference{} Seitzer, P., \& Schweizer, F. 1990, PASP, 102, 615
\reference{} Sigurdsson, S., Hernquist, L., Quinlan, G. 1995, ApJ, 446, 75
\reference{} Tormen, G. 1997, MNRAS, 290, 729
\reference{} Tremaine, S. AJ, 1995, 110, 628
\reference{} Whitmore, B. 1997, IAUS, 186, 74

\end{references}
